%% file: main.tex
\documentclass[aps,prb,twocolumn,superscriptaddress,amsmath,nofootinbib,10pt,floatfix]{revtex4-2}
\usepackage{graphicx}
\usepackage{ulem}   
\usepackage{xcolor}
\usepackage{textgreek}
\usepackage[colorlinks, linkcolor=blue, citecolor=blue]{hyperref}

\newcommand{\be}{\begin{equation}}
\newcommand{\ee}{\end{equation}}
\newcommand{\beq}{\begin{eqnarray}}
\newcommand{\eeq}{\end{eqnarray}}
\newcommand{\ba}{\[\begin{aligned}}
\newcommand{\ea}{\end{aligned}\]}

\newcommand{\la}{\langle}
\newcommand{\ra}{\rangle}

\newcommand{\Tr}{{\rm Tr\,}}

\renewcommand{\vec}[1]{{\bf #1}}
\renewcommand{\phi}{\varphi}
\renewcommand{\epsilon}{\varepsilon}
\renewcommand{\dag}{\dagger}
\def\nn{\nonumber}

\newcommand{\addJH}[1]{\textcolor{orange}{#1}}

\def \br{{\bf r}}

\def \bq{{\bf q}}
\def \bk{{\bf k}}

\def \ve{{\varepsilon}}

\def \r{{\bf {r}}}

\def \ket#1{{\,|\,#1\,\rangle\,}}
\def \bra#1{{\,\langle\,#1\,|\,}}
\def \braket#1#2{{\,\langle\,#1\,|\,#2\,\rangle\,}}

\def \ra{{\rangle}}
\def \la{{\langle}}
\def \tn{\textnormal}

\def \ba{\begin{align*}}
\def \ea{\end{align*}}

\newcounter{indice}

\begin{document}

\title{Superconductivity, charge density wave, and supersolidity in \\ flat bands with tunable quantum metric}
\author{Johannes S. Hofmann}
\author{Erez Berg}
\email{erez.berg@weizmann.ac.il}
\affiliation{Department of Condensed Matter Physics, Weizmann Institute of Science, Rehovot, 76100, Israel.}
\author{Debanjan Chowdhury}
\email{debanjanchowdhury@cornell.edu}
\affiliation{Department of Physics, Cornell University, Ithaca, New York 14853, USA.}

\begin{abstract}
Predicting the fate of an interacting  system in the limit where the electronic bandwidth is quenched is often highly non-trivial. The complex interplay between interactions and quantum fluctuations driven by the band geometry can drive competition between various ground states, such as charge density wave order and superconductivity. In this work, we study an electronic model of topologically-trivial flat bands with a continuously tunable Fubini-Study metric in the presence of on-site attraction and nearest-neighbor repulsion, using numerically exact quantum Monte Carlo simulations. 
By varying the electron filling and the minimal spatial extent of the localized flat-band Wannier wavefunctions, we obtain a number of intertwined orders. These include a phase with coexisting charge density wave (CDW) order and superconductivity, i.e., a supersolid. In spite of the non-perturbative nature of the problem, we identify an analytically tractable limit associated with a `small' spatial extent of the Wannier functions and derive a low-energy effective Hamiltonian that can well describe our numerical results.  
We also provide unambiguous evidence for the violation of any putative {\it lower} bound on the zero-temperature superfluid stiffness in geometrically non-trivial flat bands.
\end{abstract}

\maketitle

{\it Introduction.-} 
Superconductivity in narrow-band systems has attracted enormous attention, triggered in part by the discovery of two-dimensional moir\'e materials \cite{moire_rev} and the fundamental theoretical aspects that remain poorly understood \cite{Bernevig_review}. 
The limit of flat bands is particularly interesting as a possible route to optimize the superconducting $T_c$ (in the presence of an effective attraction), because of the diverging density of states. In this situation, it has been predicted that $T_c$ is proportional to $|U|$, the strength of the effective attractive interaction~\cite{Shaginyan1990,Volovik2011,Volovik2013}. However, the lack of electronic dispersion also implies a reduced superconducting phase stiffness, limiting $T_c$. Moreover, a plethora of competing non-superconducting phases may arise.

Topological flat bands, that do not admit an exponentially localized basis in real space \cite{VanderbiltRMP}, have been proposed to generate a non-zero phase stiffness within Bardeen-Cooper-Schrieffer (BCS) mean-field theory ~\cite{Torma15}.
Numerically exact determinant quantum Monte Carlo (DQMC)~\cite{Blankenbecler1981} calculations have indeed provided an unambiguous and non-perturbative demonstration for $T_c\propto|U|$~\cite{JHEBDC,Huber21,Zhang2021,Bernevig21} for topological flat-bands, where $|U|$ denotes the strength of an on-site attraction. Even for this simplified problem, the normal state for temperature $T>T_c$ exhibits strong nearly degenerate density and pairing fluctuations due to an emergent SU(2) symmetry~\cite{Torma16b}, such that the ground state is highly susceptible to competing orders. Within BCS mean-field theory and for models satisfying a set of restrictive conditions, lower bounds on the zero-temperature phase stiffness have been proposed ~\cite{Torma15,Bernevig19,Rossi19,Torma19}, and shown to be governed by the integrated Fubini-Study metric (up to an energy-scale set by the superconducting gap). Beyond mean-field theory, upper bounds on the stiffness have also been proven ~\cite{Hazra18,MR21,DM21}.

This naturally leads to the following questions when departing from the BCS paradigm:
(i) What is the nature of the competing phases and associated quantum phase transitions that arise in flat bands as a function of various microscopic tuning parameters?  (ii) How does varying the minimal spatial extent of the localized Wannier functions tune the system between different orders? (iii) Is there a theoretical limit in which this band competition can be explored in a controlled fashion without resorting to uncontrolled mean-field theory? (iv) Can infinitesimal  perturbations drive competing instabilities leading to substantial violations of proposed lower bounds on $T_c$? 

\begin{figure}
	\begin{center}
		\includegraphics[width=0.99\columnwidth]{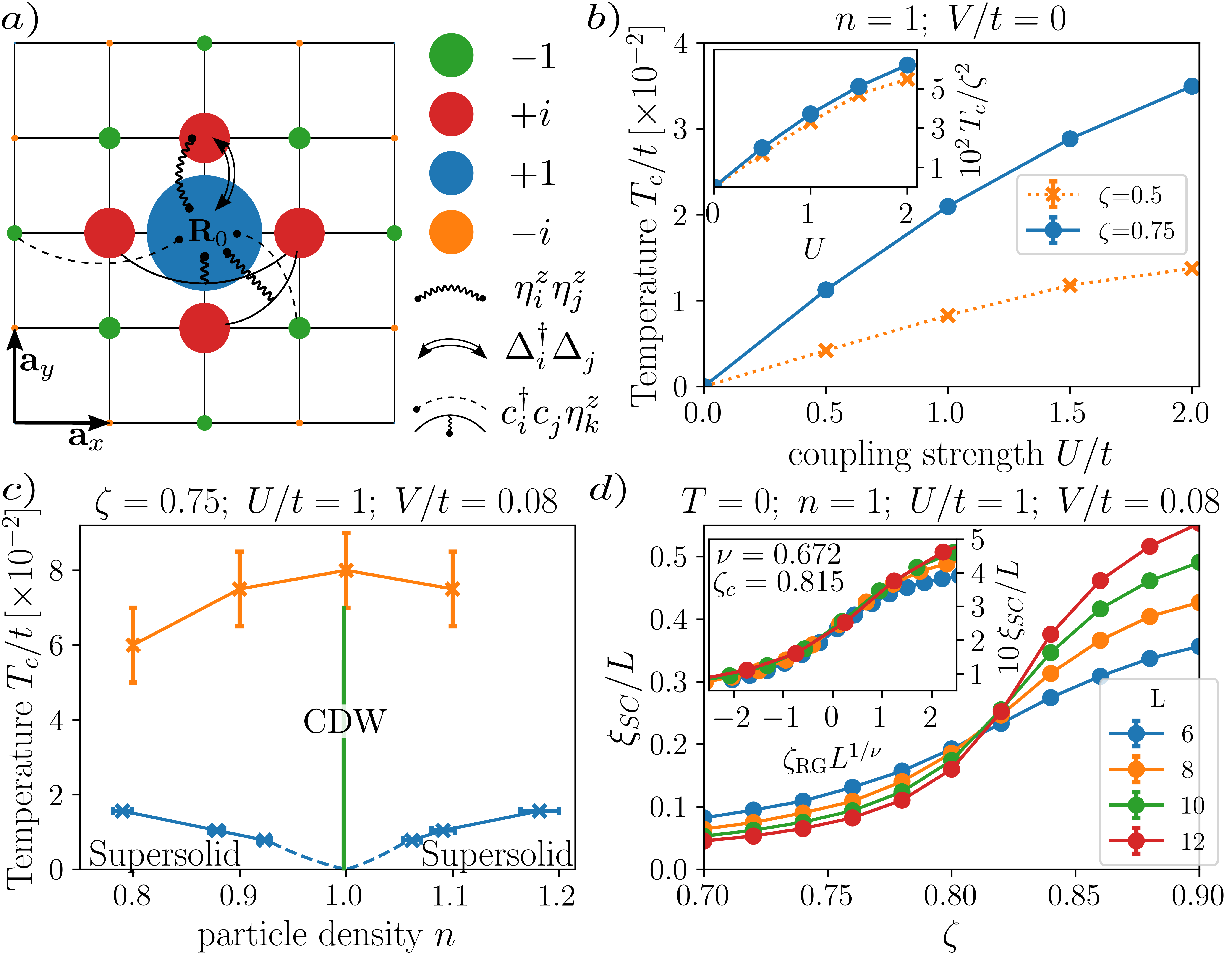}
	\end{center}
	\caption{ a) The localized Wannier function, $\Phi_{\vec{R}_0}({\vec{r}})$. The area (color) of the disc is proportional to the amplitude (phase). 
		b) The superconducting $T_c$ increases with $\zeta^2|U|$ (see also inset). 
		c) The phase diagram for $V=0.08$.
		At $n=1$, the system is an insulating CDW for $T \lesssim 0.075$, and when doped, the excess carriers lead to a supersolid. 
		d) The normalized superconducting correlation length $\xi_{\mathrm{SC}}/L$ at $T=0$ across a CDW to supersolid transition with increasing $\zeta$, consistent with a $(2+1)-$dimensional XY phase transition ($\zeta_c=0.815$, $\zeta_{\mathrm{RG}}=(\zeta-\zeta_c)/\zeta_c$) for different system sizes, $L$ .
	}
	\label{fig:sum}
\end{figure}

In this Letter, we study a concrete electronic Hamiltonian for topologically trivial flat bands, where the minimal spatial extent of the exponentially localizable Wannier functions can be tuned {\it continuously} without affecting the band dispersion. Using unbiased DQMC simulations, we will demonstrate that such a model supports superconductivity with a wide fluctuation regime and obtain the detailed dependence of $T_c$ on the spatial extent of the Wannier functions. 
Additionally, by varying the strength of further-neighbor interactions, the electronic density, and the spatial extent of the Wannier functions, we can drive continuous phase transitions to
charge density wave (CDW) and supersolid phases within the same flat-band limit. Remarkably, we can use the spatial extent of the Wannier functions as a `small' parameter to derive an effective pseudospin Hamiltonian that helps explain the intertwined superconductivity, CDW, and supersolid orders in this flat-band model. Finally, the ability to tune continuously between various (non-) superconducting phases allows us to violate any putative lower bound on $T_c$.

{\it Model.-} We define a two-orbital, spinful electronic model with local interactions. The model, first introduced in Ref.~\cite{boundless}, is time-reversal symmetric, and resides on a square two-dimensional lattice with two orbitals per site. We will focus on densities in the vicinity of one electron per unit cell, corresponding to quarter filling. The model exhibits two pairs of perfectly flat bands at energies, $\varepsilon_\bk=\pm t$, where $t$ sets the overall scale associated with the microscopic hopping parameters. The energy gap is $\Delta_{\tn{gap}}=2t$. 

The non-interacting part of the Hamiltonian reads
\beq\label{eq:Ham0}
H_0 &=& -t \sum_{\bk} \hat{\mathbf{c}}^{\dagger}_{\bk} \left(\lambda_x \sin{\alpha_{\bk}} + \sigma_z \lambda_y \cos{\alpha_{\bk}} + \mu \lambda_0 \right) \hat{\mathbf{c}}^{\phantom{\dagger}}_{\bk},\nn \\
\alpha_{\bk}&=&\zeta (\cos{k_x a}+\cos{k_y a}).
\eeq
Here, $\hat{\mathbf{c}}^{\dagger}_{\bk}$ is a vector of operators $\hat{c}^{\dagger}_{\bk,l,s}$ which create electrons with momentum $\bk$ and spin $s=\uparrow,\downarrow$ in orbital $l=1,2$. The Pauli-matrices $\sigma_{j=0,x,y,z}$ and $\lambda_{j=0,x,y,z}$ act on the spin and orbital indices, respectively, and $a=|\mathbf{a}_x|$ is the length of the primitive lattice vectors.
Regardless of $t$, the dimensionless parameter $\zeta$ controls the spatial extent of the localized Wannier functions, $\Phi_{{\bf R}_0}({\bf r})\sim (i\zeta)^{|\delta_x|+|\delta_y|} + \mathcal{O}(\zeta^{|\delta_x|+|\delta_y|+2})$,
where $\textbf{\textdelta} = (\r-{\bf R}_0)/a$; see Fig.~\ref{fig:sum}(a). The quantum geometric tensor, ${\cal{G}}_{ij}$, is simple---the imaginary part (i.e. Berry curvature) vanishes everywhere in the BZ while the real part (i.e. the Fubini-Study metric) is finite and integrates to $\zeta^2 a^2/2$ \cite{si}. 
Note that the metric depends on how the orbitals are embedded in real space; here, both orbitals are located at the center of the unit cell in the $x-y$ plane, respecting $C_4$ rotation symmetry \cite{Simon2020}.

At a fractional filling of the lower band ($\ve_{\bf k}=-t$), we will study the effect of on-site attraction, $U>0$, and nearest-neighbor interaction, $V$, 
\beq
\label{eq:HamInt}
H_{\tn{int}} &=& - \frac{U}{2} \sum_{\r,l}  \delta \hat{n}_{\r,l}^2
    + V \sum_{\langle \r,\r'\rangle,l} \delta \hat{n}_{\r,l} ~\delta \hat{n}_{\r',l} \,, \label{chiral}
\eeq
where $\delta \hat{n}_{\r,l} = \sum_s \hat{c}^{\dagger}_{\r,l,s}\hat{c}^{\phantom{\dagger}}_{\r,l,s} - 1$ refers to the (shifted) density operator in orbital $l$ at site $\r$.  The above model, $H=H_0 + H_{\tn{int}}$, is free of the infamous sign problem as long as $U\ge 4|V|$. Before analyzing the model numerically,
%\addEB{\sout{at a finite temperature $T$}}, 
we derive the effective Hamiltonian that illustrates the competition between various ordering tendencies. This analytical approach relies on a controlled expansion for small $\zeta$,  but agrees qualitatively with the non-perturbative results obtained using DQMC.

{\it Analytical results for small $\zeta$.--} We focus on the limit of $\zeta \ll 1$ and $T,V\ll U \ll\Delta_{\tn{gap}}(=2t)$, that allows us to project the interaction to the lower ``active" band. 
The localized Wannier wave function of the lower band, centered around $\r={\bf R}_0$, $\Phi_{{\bf R}_0,s}(\r) = \frac{1}{\sqrt{2}L^2}\sum_{\bk}e^{i\bk({\bf R}_0-\r)} e^{is \lambda_z \alpha_{\bk}/2}(1,-\,s i)^\dag$~\cite{si}, is depicted in Fig.~\ref{fig:sum}(a). Upon introducing new operators, $\hat{c}^{\dagger}_{\r,l,s} \mapsto \sum_{\r'} \Phi^*_{\r,s}(\r',l) \hat{d}^{\dagger}_{\r',s}$, the projected interaction Hamiltonian in the $\zeta\ll1$ limit takes the form of an effective XXZ model supplemented by other terms,
\begin{eqnarray}
	\widetilde{H_{\tn{int}}} &=& - \frac{U_{\tn{eff}}}{2} \sum_{\r} \left(2\hat{\eta}^z_{\r}\right)^2  + \frac{U\zeta^2}{32}\sum_{\r}\hat{\eta}^z_{\r}(2 \hat{B}^{\delta}_{\r} - \hat{B}^{2\delta}_{\r} )   \nonumber \\
	&& -\sum_{\langle \r, \r'\rangle} [J_\perp(\hat{\eta}^x_{\r} \hat{\eta}^x_{\r'} + \hat{\eta}^y_{\r} \hat{\eta}^y_{\r'}) + J_z \hat{\eta}^z_{\r} \hat{\eta}^z_{\r'} ]  \,,\label{eq:effHam}
\end{eqnarray}
with pseudospin operators, $\hat{\eta}^{j=0,x,y,z}_{\r} \equiv (\Psi^\dagger_\r \eta^j \Psi^{\phantom{\dagger}}_\r)/2$, where $\Psi^\dagger_\r = (\hat{d}^{\dagger}_{\r,\uparrow}, \hat{d}^{\phantom{\dagger}}_{\r,\downarrow})$ and $\eta^j$ are Pauli matrices.

The parameters, $U_{\tn{eff}}=U(2-\zeta^2)/4$, $J_\perp=\zeta^2 U/4$, and $J_z=\zeta^2 U /4 - 2V$. For $\zeta=0$, the sites decouple completely and $\Phi_{\r,s}(\r')\propto\delta_{\r \r'}$; only the first term in Eq.~\eqref{eq:effHam} survives and the ground state manifold is highly degenerate, consisting of local Cooper pairs without long-ranged phase coherence. The projected Hamiltonian also contains interaction-mediated nearest-neighbor, $\hat{B}^{\delta}_{\r} = \sum_{\substack{s\addJH{,} \mathbf{e}_1,\mathbf{e}_2\\\mathbf{e}_1\neq \mathbf{e}_2}} \hat{d}^{\dagger}_{\r+\mathbf{e}_1,s}\hat{d}^{\phantom{\dagger}}_{\r+\mathbf{e}_2,s}$, and second nearest-neighbor, $\hat{B}^{2\delta}_{\r} = \sum_{\substack{s\addJH{,} \mathbf{e}_1,\mathbf{e}_2\\\mathbf{e}_1\neq -\mathbf{e}_2}} (\hat{d}^{\dagger}_{\r+\mathbf{e}_1+\mathbf{e}_2,s}\hat{d}^{\phantom{\dagger}}_{\r,s} + \tn{h.c.} )$, hopping terms, with $\mathbf{e}_{1,2} \in \{\pm\mathbf{a}_x,\pm\mathbf{a}_y\}$. 
In Fig.~\ref{fig:sum}(a), we represent the pair hopping ($J_\perp$) and nearest-neighbor density ($J_z$) interactions by double-solid and wiggly lines, respectively. The interaction-mediated hoppings $\hat{B}^\delta$ and $\hat{B}^{2\delta}$ are depicted as solid and dashed lines, respectively. The interaction-mediated hopping between nearest-neighbor sites at order $\zeta$ vanishes due to 
chiral symmetry~\cite{si}.

At finite $\zeta$ and $V=0$, $\widetilde{H_{\tn{int}}}$ exhibits an emergent SU(2) symmetry~\cite{Torma16b} and strong fluctuations in the degenerate density and pairing response, without any long-range order at finite temperature. This symmetry is broken by higher-order terms in $U/\Delta_{\tn{gap}}$, leading to an anisotropy $\Delta J= J_\perp-J_z$ and a finite superconducting transition temperature with $T_c\propto \pi J_\perp/\log(\pi J_\perp/\Delta J)$, as shown in Fig.~\ref{fig:sum}(b).
The anisotropy can also be tuned by turning on $V$; the ground state is susceptible towards formation of an ordered CDW at a commensurate filling  when $-|J_\perp|>J_z$ (Fig.~\ref{fig:sum}c). Doping away from the commensurate CDW at $n=1$ induces a density-mediated hopping and leads to a supersolid phase with long-range superconducting phase coherence (Fig.~\ref{fig:sum}c). Furthermore, at $n=1$, increasing $\zeta$ also induces a continuous transition to a supersolid ground state, consistent with the $(2+1)$-dimensional XY universality class (Fig.~\ref{fig:sum}d-inset).

\begin{figure}
	\begin{center}
		\includegraphics[width=0.99\columnwidth]{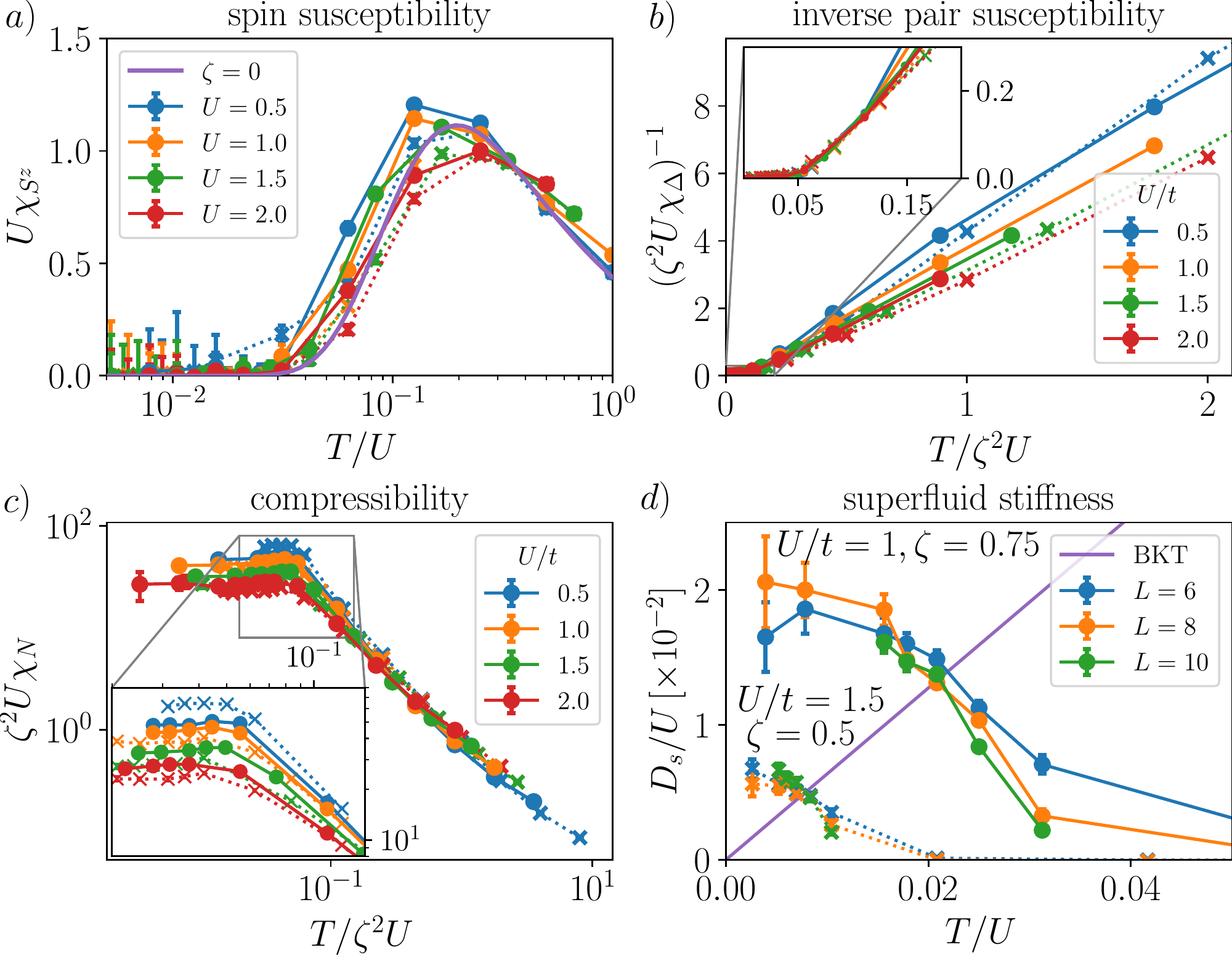} 
	\end{center}
	\caption{\label{fig:Tc_vs_U} (a) Spin susceptibility, $\chi_{S^z}(T)$, for different $U$ at fixed $\zeta=0.75$ (solid) and $\zeta=0.5$ (dashed). The purple line, $\zeta=0$, represents the atomic limit with a spin gap of $\Delta_{S^z}=U/4$. (b) Inverse pair susceptibility, $\chi_\Delta^{-1}(T)$, and (c) compressibility, $\chi_N(T)$, as a function of temperature obtained for same values of $U$ as in (a). (d) Exemplary data for superfluid stiffness, $D_s(T)$, used to extract the critical temperature, $T_c$. Results obtained for $V=0$.
	}
\end{figure}

{\it Numerical results.-} We note that $H_0$ contains hopping matrix elements which decay in real-space as $t_{\textbf{\textdelta}}\sim \zeta^{|\delta_x|+|\delta_y|}$. We truncate the range of hopping in our implementation of the DQMC computations using ALF~\cite{ALF2017,ALF2021}, neglecting terms with $|\delta_x|+|\delta_y|>3$, leading to non-zero bandwidth $W\sim O(\zeta^4)$.

We first focus on the case of on-site interaction only ($V=0$) at quarter filling ($n=1$). We are interested in two-particle susceptibilities of local operators, $\hat{O}$, i.e. $\chi_{O}=L^{-2} \int_0^\beta d\tau \la \hat{O}(\tau)\hat{O}(\tau=0) \ra$ with inverse temperature $\beta=(k_B T)^{-1}$ and imaginary time $\tau$.
For instance, for $\hat{O}\equiv S_z$, $\chi_{S^z}$ is the spin susceptibility. The results for $\chi_{S^z}$ vs. temperature are shown in Fig.~\ref{fig:Tc_vs_U}(a) for few different interaction strengths and $\zeta=0.5$ (dashed line) or $\zeta=0.75$ (solid line). The data obey nearly perfect scaling of the form $\chi = f(T/U)/U$.  $\chi_{S^z}$ is peaked near $T\sim0.2U$ and shows a dramatic suppression for $T\lesssim0.1U$. The onset of such a ``pseudogap" behavior is already present in the $\zeta\rightarrow0$ limit (purple curve in Fig.~\ref{fig:Tc_vs_U}a), where the gap $\Delta_{S^z} = U/4$. 

In addition, we examine the pairing-susceptibility, $\chi_\Delta$ for $\hat{O}\equiv \Delta_s = \sum_{\r,l} (c_{\r,l,\uparrow} c_{\r,l,\downarrow} + \tn{h.c.})$, and the charge-compressibility, $\chi_N$ for $\hat{O}\equiv N = \sum_{\r,l,s} (c^{\dag}_{\r,l,s}c_{\r,l,s}-n)$. The pairing and charge fluctuations are strongly enhanced with decreasing temperature, signaling a near degeneracy between the competing tendencies towards superconductivity and phase-separation \cite{Torma16b}; see Fig.~\ref{fig:Tc_vs_U}(b)-(c).
However, upon approaching the superconducting $T_c$ from above, the pair-susceptibility diverges (i.e., $\chi_\Delta^{-1}\rightarrow0$), while the compressibility saturates to a finite value.

In two dimensions, the superconducting $T_c$ can be obtained using the criterion $D_s(T\rightarrow T_c^-)=2T_c/\pi$~\cite{NK77}, where $D_s(T) = -[K_x + \Lambda_{xx}(\bq\rightarrow0)]/4$ is the superfluid stiffness; $\Lambda_{xx}(\bq)$ is the paramagnetic current-current correlation function at zero Matsubara frequency, and $K_x$ is the diamagnetic current correlator~\cite{SFcriteria}. In Fig.~\ref{fig:Tc_vs_U}(d), we show the data for $D_s(T)$ for $(U,\zeta)=(1.0,0.75)$ (solid) and $(U,\zeta)=(1.5,0.5)$ (dashed). To a reasonable approximation, $T_c\propto U\zeta^2$, as shown in Fig.~\ref{fig:sum}(b). This is expected based on our discussion of the effective XXZ model in the small $\zeta$ limit.  A superconducting instability with $T_c\propto U$ has been reported in earlier DQMC computations involving topological flat-bands \cite{JHEBDC,Huber21},  
and more recently in topologically {\it trivial} flat-bands \cite{Bernevig21}. Our numerically exact analysis of this non-perturbative regime and the complementary analytical results obtained using the XXZ pseudospin Hamiltonian offer new insights into the role of a tunable metric in flat-band superconductors. 

We now include a repulsive nearest-neighbor density interaction, $V=0.08$, and analyze the phase diagram for a range of fillings near $n=1$ (Fig.~\ref{fig:sum}c). The main effect of the repulsive interaction is to spontaneously break the discrete translational symmetry and induce a CDW order at an ordering wavevector of $(\pi,\pi)$. 
We have extracted the CDW correlation length~\cite{Toldin15,si}, $\xi_{\tn{CDW}}$, and the associated transition temperature $T_{\tn{CDW}}$ as a function of $n$ for a range of fillings near $n=1$ (SI Sec. E \cite{si}); note that a fully insulating CDW is present only at the commensurate filling $n=1$ (green vertical line in Fig.~\ref{fig:sum}c). Next, we address the fate of this insulating CDW when doped with electrons or holes away from $n=1$.

\begin{figure*}
	\begin{center}
		\includegraphics[width=\textwidth]{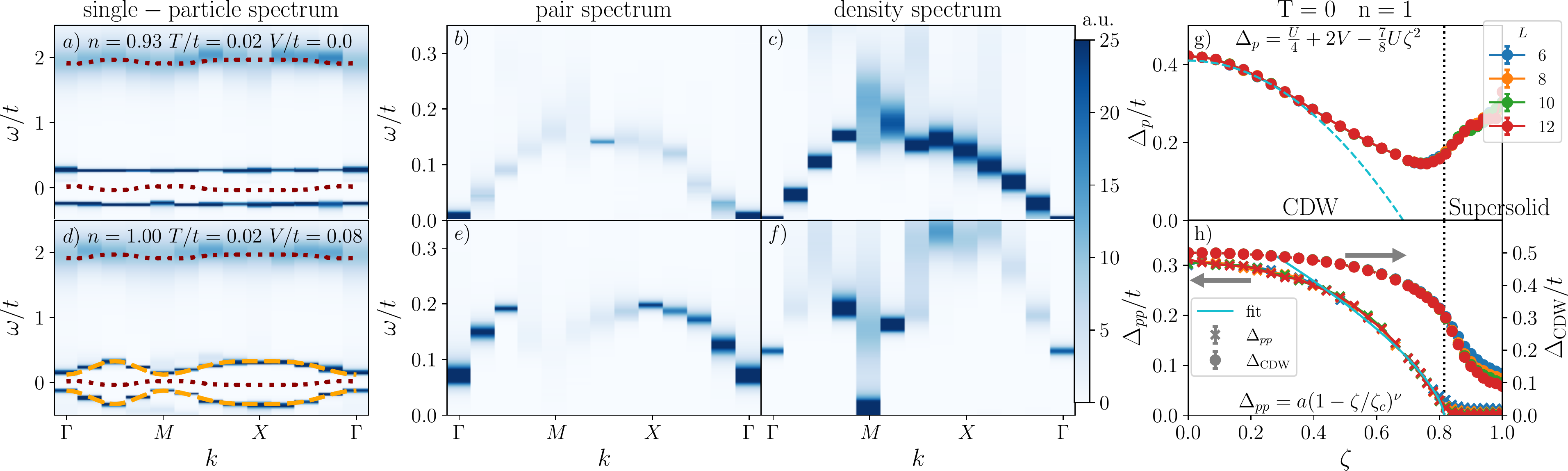}
	\end{center}
	\caption{\label{fig:dynamics} Excitation spectra
	in the superconducting phase (a-c), in the CDW phase (d-f) at $T/t=0.02$ [$\zeta=0.75$], and excitation energies relative to the ground state (g-h):
	(a) and (d) show electronic spectra featuring the flat and interaction-induced dispersive bands (non-interacting bands in maroon). (b) and (e) show s-wave pair spectra. (c) and (f) show density spectra. (g) and (h) display the single- and two-particle excitation energies, respectively, of the ground state at quarter filling with $U=1$ and $V=0.08$. The turquoise fits depict the expectations from the density-assisted hopping (g) and 3D-XY universality (h).
	}
\end{figure*}

Furthermore, we analyze $D_s$ at low temperature, $T=0.008$, as a function of particle density (SI Sec. E \cite{si}).
We identify BKT transitions towards superconducting order and find the critical carrier densities
$n_c=0.923\pm0.006$ and $n_c=1.062\pm0.008$ for hole and electron doping, respectively. 
Fig.~\ref{fig:sum}(c) summarizes $n_c$ for different temperatures. $T_c$ vanishes for $n=1$ and increases monotonically with $\delta n=|n-1|$, suggesting that superconductivity arises due to excess (``doped") electrons or holes relative to the ordered CDW insulator. 
Importantly, along with superconductivity, the CDW remains long-range ordered. Thus the resulting phase is a supersolid, with a finite superconducting phase stiffness and a spontaneously broken lattice translational symmetry. The lightly doped system can effectively be described in terms of a dilute liquid of interacting bosons with a superconducting $T_c\propto \delta n$, up to additional logarithmic corrections \cite{popov1972theory,kagan1987influence,Fisher1988}.

We also examine the single and two-particle spectrum in the different phases \cite{si}. 
We compute $A(\bk,\omega)=-\textrm{Im}\, G(\bk,\omega)/\pi$ from the imaginary time Green's function, $G(\bk,\tau)=\sum_{l,s} \la c^{\phantom{\dag}}_{\bk,l,s}(\tau)c^\dag_{\bk,l,s}(0) \ra$, and the pairing and density spectra via the maximum entropy method ~\cite{MaxEnt2004}. For $V=0$, the resulting single and two-particle spectra are summarized in Figs.~\ref{fig:dynamics}(a)-(c).
The single electron spectrum 
exhibits two nearly flat bands at $\omega\approx\pm 0.25 t$, that are well separated from a broader band at $\omega\approx2 t(=\Delta_{\tn{gap}})$. The latter is clearly the higher energy flat band associated with $H_0$. The bands at $\omega\approx\pm 0.25 t$ originate from the low-energy flat band of $H_0$, that splits due to the Hubbard interaction; specifically, the splitting energy is approximately $\Delta_{S^z}=U/4$, the gap associated with the $\zeta=0$ limit. The spectrum of pairing and density excitations show linearly dispersing, Goldstone-like modes near the $\Gamma-$point. These modes can be understood as arising from the approximate SU(2) symmetry of the attractive Hubbard interaction projected to the flat band~\cite{Torma16a}. 

For $V=0.08$, the single and two-particle spectra are summarized in Figs.~\ref{fig:dynamics}(d)-(f).
The high-energy band at $\omega\approx2 t(=\Delta_{\tn{gap}})$ in Fig.~\ref{fig:dynamics}(d) is nearly identical to the previous case. However, the low-energy bands are significantly more dispersive than in Fig.~\ref{fig:dynamics}(a) due to the density-assisted hopping terms (Fig.~\ref{fig:sum}) of the projected Hamiltonian.
Note that adding a single electron to the background of the CDW, on the one hand, costs an energy $\bar{\Delta}$ due to breaking a pair and creating a point defect in the CDW; this accounts for the finite energy offset and, in the small $\zeta$ limit, this energy is  $\bar{\Delta}=\frac{U_{\tn{eff}}}{2}-4\Delta^2_{\tn{CDW}}J_z$~\cite{si}. On the other hand, the electron can delocalize and gain kinetic energy due to the density-assisted hopping where the effective width of the shifted band scales as $U \zeta^2 \Delta_{\tn{CDW}}$ with $\Delta_{\tn{CDW}}\equiv\la e^{i\r\bq_{\mathrm{CDW}}} \hat{\eta}^z_\r \ra$ and $\bq_{\mathrm{CDW}}=(\pi,\pi)$. The dispersion relation directly follows from the second term of Eq.~\eqref{eq:effHam} and is depicted as dashed orange lines in Fig.~\ref{fig:dynamics}(d).

The two-particle spectra show that the linearly dispersing Goldstone mode near the $\Gamma-$point is gapped for $n=1$ and $V=0.08$, while the density spectrum (Fig.~\ref{fig:dynamics}f) exhibits a clear softening near the CDW ordering wavevector. Increasing $\zeta$ while keeping all other parameters fixed, we have extracted the single-particle and two-particle gaps, $\Delta_p$ and $\Delta_{pp}$, at $T=0$ near the $\Gamma-$point (Figs.~\ref{fig:dynamics}g-h, respectively). For $\zeta<0.7$,  $\Delta_p$ decreases with increasing $\zeta$, in agreement with the expectation $\Delta_p=\frac{U}{4} + 2V-\frac{7}{8}U\zeta^2$ \cite{si}. $\Delta_p$ assumes its minimal value at $\zeta\approx 0.75$. Similarly, $\Delta_{pp}$ is largest for $\zeta=0$ and decreases with increasing $\zeta$ (Fig.~\ref{fig:dynamics} h), vanishing for $\zeta > 0.815$. It is important to note that the onset of superconductivity at $\zeta_c=0.815$, as inferred from the behavior of  $\xi_{\mathrm{SC}}/L$ (Fig.~\ref{fig:sum}d), is accompanied by a finite $\Delta_{\tn{CDW}}$; the transition from CDW to supersolid order belongs in the $(2+1)$-dimensional XY universality class.
In the pseudospin notation introduced in Eq.~\ref{eq:effHam}, the supersolid is represented by a canted antiferromagnet, where the in-plane (XY) ferromagnetic components represent the SC while the out-of-plane antiferromagnetic component represents the CDW.

{\it Discussion.-} Our work highlights phase competition in flat bands with vanishing Berry curvature but non-trivial Fubini-Study metric. By construction, these systems are strongly correlated; in addition, due to the band geometry, quantum fluctuations are important even in the perfectly flat band limit. As a result, the phase diagram can be difficult to predict 
{a priori}, without controlled calculations. 

We have demonstrated this by a sign problem-free, explicitly solvable model with a tunable quantum metric. The model exhibits a cascade of quantum phases. The interactions within the flat band lead to the formation of a CDW phase, whose electronic excitations acquire a non-trivial dispersion due to the band geometry. Increasing this dispersion by tuning the quantum metric ultimately leads to a further instability towards a supersolid phase. 

We expect such cascades of different ordering tendencies to arise also in realistic flat-band systems, such as those that occur in two-dimensional van der Waals materials.
Additionally, there is a promising prospect to engineer and directly simulate some elements of the models considered here in future cold-atoms based experiments. Recent experiments using ultracold bosonic atoms have identified supersolids in one \cite{Li17_supersolid_exp,Leonard17_supersolid_exp,Tanzi19_supersolid_exp,Guo19_supersolid_exp,Natale2019_supersolid_exp} and two dimensions \cite{Norcia2021_supersolid_2D,Biagioni2022_supersolid_2D}.  Realizing supersolids in models of fermionic ultracold atoms \cite{Bloch22_overview,Esslinger2010_fermion_QSim_rev,Vale2021_QSim_rev}  remains an interesting open challenge, but can potentially be realized using the setups proposed here. 

Furthermore, our work unambiguously demonstrates that any proposed lower bound on the superfluid stiffness in terms of single-properties of the flat band, such as the quantum geometry, are strictly inapplicable beyond BCS mean-field theory. In the presence of a large on-site attraction and weak nearest-neighbor repulsion, where application of the mean-field approximation will lead one to conclude a superconducting ground state with a non-zero superfluid stiffness, our exact computations show that the stiffness can be made arbitrarily small and even vanish, violating any putative bound.

{\it Acknowledgements.-} The authors thank Ankur Das, Tobias Holder, and Steven Kivelson for stimulating discussions. DC is supported by faculty startup funds at Cornell University. 
EB and JH were supported by the European Research Council (ERC) under grant HQMAT (grant no. 817799), the US-Israel Binational Science Foundation (BSF), and a Research grant from Irving and Cherna Moskowitz. 
This work used the Extreme Science and Engineering Discovery Environment (XSEDE), which is supported by National Science Foundation grant number ACI-1548562~\cite{xsede}. The authors gratefully acknowledge the generous computing time on Expanse at the San Diego Supercomputer Center through allocation TG-PHY210006. We also thank Cyrus Umrigar for sharing his computational resources (TG-PHY170037) at the early stages of this work. 
The auxiliary field QMC simulations were carried out using the ALF package available at \url{https://alf.physik.uni-wuerzburg.de}.

\bibliography{references}

% Comment out the following lines for APS submission
\clearpage
\input{supp-core}

\end{document}

%% file: supp-core.tex
\renewcommand{\thefigure}{S\arabic{figure}}
\renewcommand{\figurename}{Supplemental Figure}
\setcounter{figure}{0}
\begin{widetext}
\appendix
{\bf \centering SUPPLEMENTARY INFORMATION for\\``Superconductivity, charge-density wave and supersolidity in flat-bands with tunable quantum metric"}

\section{Wannier wave function}
The non-interacting Hamiltonian, Eq.~(1) of the main text, can be rewritten in a rotating frame as
\begin{eqnarray}
	H_s(\mathbf{k}) &=& -t \left( \lambda_x\sin{\alpha_{\bk}} + (-1)^s \lambda_y \cos{\alpha_{\bk}}\right)\nn \\
	    &=& -t (-1)^s e^{i \frac{(-1)^s \alpha_{\bk}}{2}  \lambda_z}  \lambda_y e^{-i \frac{(-1)^s \alpha_{\bk}}{2}  \lambda_z}  \,.
	    \label{eq:Ham0pauliRot}
\end{eqnarray}
Then the eigenstates at momentum $\mathbf{k}$, satisfying 
\begin{equation}
H_s(\mathbf{k}) \phi_{\pm,{\bf k},s} = \pm t\, \phi_{\pm,{\bf k},s}
\end{equation}
are readily given by the rotating eigenvectors of $\lambda_y$, 
\be
\phi_{\pm,{\bf k},s} = 1/\sqrt{2} e^{i \frac{(-1)^s \alpha_{\bk}}{2}  \lambda_z} \left(1,\mp (-1)^s i \right)^T\,.\label{eq:wavefunction}
\ee

To derive the exponentially localized Wannier orbitals that span the lower energy band, we introduce,
\begin{equation}
 \hat{ \mathbf{c} }^{\dagger}_{\br_i,s} = \frac{1}{\sqrt{A}} \sum_{\bk} e^{-i\br_i\cdot\bk} \, \hat{ \mathbf{c} }^{\dagger}_{\bk,s},
 \quad
 \hat{ \mathbf{c} }^{\dagger}_{\bk,s} = \frac{1}{\sqrt{A}} \sum_{\br_i} e^{i\br_i\cdot\bk} \, \hat{ \mathbf{c} }^{\dagger}_{\br_i,s}
 \,.
\end{equation}
Hence, we have,
\begin{eqnarray}
    \hat{ c }^{\dagger}_{\br_i,l,s} & = & 
    \frac{1}{\sqrt{A}} \sum_{\bk} e^{-i\br_i\bk} \, \hat{ c }^{\dagger}_{\bk,l,s} =
    \frac{1}{\sqrt{A}} \sum_{\bk} e^{-i\br_i\bk} \sum_{\lambda=\pm} \phi^*_{\lambda,\bk,s}(l) \hat{ d }^{\dagger}_{\bk,\lambda,s} \nonumber \\
    &\rightarrow & 
    \frac{1}{\sqrt{A}} \sum_{\bk} e^{-i\br_i\bk} \phi^*_{-,\bk,s}(l) \hat{ d }^{\dagger}_{\bk,-,s} =
    \frac{1}{A} \sum_{\bk,\br_j} e^{-i\br_i\bk} \phi^*_{-,\bk,s}(l) e^{i\br_j\bk} \, \hat{ d }^{\dagger}_{\br_j,-,s} = \sum_{\br_j} \Phi^*_{\br_i,s}(\br_j,l) \, \hat{ d }^{\dagger}_{\br_j,s} \label{eq:c-projected}\\
    \Phi_{\br_i,s}(\br_j,l) &=& \frac{1}{A} \sum_{\bk} e^{i(\br_i-\br_j)\bk} \phi_{-,\bk,s}(l)\,,
\end{eqnarray}
where we dropped the band index in the last step of Eq.~\eqref{eq:c-projected}. The Wannier orbitals are given by $\Phi_{\br_i,s}(\br_j,l)$ and $\hat{ d }^{\dagger}_{\br_i,s}=\frac{1}{\sqrt{A}} \sum_{\bk} e^{-i\br_i\bk} \, \hat{ d }^{\dagger}_{\bk,-,s}$ creates an electron centered around the $\br_i^{\mathrm{th}}$ unit cell with spin $s$.

\section{Quantum geometric tensor}

The quantum geometric tensor of the lower Bloch band, $u_{-,{\bf k},s}$, is given by
\be
\mathcal{G}_{ij}(\bk,s)=\bra{\partial_i u_{-,{\bf k},s}} \left(1-\ket{u_{-,{\bf k},s}}\bra{u_{-,{\bf k},s}}\right)\ket{\partial_j u_{-,{\bf k},s}}\,.\label{eq:geo_tensor}
\ee
For point-like orbitals, Bloch state $u$ and the Hamiltonian eigenstate $\phi$ are related by $u_{-,{\bf k},s,l}=e^{-i{\bf k  x}_l}\phi_{-,{\bf k},s,l}$ where ${\bf x}_l$ is the orbital position \cite{Simon2020}. Note that we have ${\bf x}_0={\bf0}$ and ${\bf x}_1=-a{\bf e}_z$, where ${\bf e}_z$ is the unit vector in the $z$ direction, i.e., the two orbitals reside at the origin of the unit cell in the $xy$ plane and are displaced in the z-direction such that $l=0$ ($l=1$) is the upper (lower) layer. Hence, we have $u_{-,{\bf k},s}=\phi_{-,{\bf k},s}$.
The derivatives are readily determined from Eq.~\eqref{eq:wavefunction}, 
\begin{eqnarray}
    \ket{\partial_j \phi_{-,{\bf k},s}} = i \left(\partial_j \alpha_{\bk}\right) \frac{(-1)^s }{2}  \lambda_z \ket{ \phi_{-,{\bf k},s}}\,,
    \quad \quad & & \quad \quad
    \bra{\partial_i \phi_{-,{\bf k},s}} = -i \left(\partial_i \alpha_{\bk}\right) \frac{(-1)^s }{2} \bra{ \phi_{-,{\bf k},s}}  \lambda_z \,.
\end{eqnarray}
Note that due to the chiral symmetry, the wave function has equal support on both orbitals such that $ \bra{ \phi_{-,{\bf k},s}} \lambda_z \ket{\phi_{-,{\bf k},s}} = 0$ and hence $\braket{ \phi_{-,{\bf k},s}}{ \partial_j \phi_{-,{\bf k},s}} = 0$. Therefore, we have
\begin{eqnarray}
    \mathcal{G}_{ij}(\bk,s) & = & \braket{\partial_i \phi_{-,{\bf k},s}} {\partial_j \phi_{-,{\bf k},s}} \\ 
    & = & \frac{1}{4}  \left(\partial_i \alpha_{\bk}\right) \left(\partial_j \alpha_{\bk}\right) \bra{ \phi_{-,{\bf k},s}}  \lambda_z  \lambda_z \ket{ \phi_{-,{\bf k},s}} \\
    & = & \frac{1}{4} \zeta^2 a^2 \sin{k_i a} \sin{k_j a}
\end{eqnarray}
and the quantum geometric tensor integrates to
\be
    \frac{a^2}{(2\pi)^2}\iint_0^{2\pi/a}dk_x dk_y \mathcal{G}_{ij}(\bk,s) = \frac{\zeta^2}{8} a^2 \delta_{ij}\,, \hspace{2cm} \frac{a^2}{(2\pi)^2}\iint_0^{2\pi/a}dk_x dk_y \sum_s \Tr [\mathcal{G}_{ij}(\bk,s)] = \frac{\zeta^2}{2} a^2\,.
\ee

\section{General remarks about the orbital embedding}
\label{app:orbital-embedding}

%We would like to begin by pointing out that 
The superfluid stiffness, a thermodynamic response function evaluated at zero external frequency and the limit of transverse momentum taken to zero, i.e., in the long-wavelength limit, does not depend on the specific choice of real-space orbital embedding for a model with a fixed electronic spectrum. However, the orbital embedding can influence other quantities, as we elaborate below (see also Ref.~\cite{Simon2020}).

The orbital embedding typically influences the hopping amplitude of the tight-binding model as it relates to the orbital overlap integral. Furthermore, the position of the orbitals may influence the point group symmetry, e.g., shifting the orbitals away from the high symmetry points would break (parts) of the $C_4$ rotation symmetry. Also, the relative position of two orbitals determines the current operator which is derived from minimal coupling, e.g., bonds that are orthogonal to the vector potential do not contribute to the electrical current.

In addition to these  aspects, the orbital embedding also plays a role when calculating the quantum geometric tensor, as indicated in the previous section. In particular, the phase factor $e^{-i{\bf k  x}_l}$ due to the orbital position contributes to derivatives in \eqref{eq:geo_tensor}.
To facilitate the discussion of how the embedding modifies the quantum geometric tensor, we follow a notation similar to Ref.~\cite{Simon2020}. 
Since the non-interacting model discussed here is block diagonal in spin, let us focus on a single spin sector as an effective two-band system, suppress the spin subscript $s$ for readability and define the generalized Berry connection $\mathbf{A}_{nm}(\bk)$ and shift vector $\mathbf{\delta x}_{nm}(\bk)$ as
\begin{eqnarray}
    \mathbf{A}_{nm}(\bk) &=& i\langle u_{n,\bk}|\nabla_\bk|u_{m,\bk}\rangle \\
    \mathbf{\delta x}_{nm}(\bk) &=& \sum_l \Delta \mathbf{x}_l u_{n,\bk,l}^* u_{m,\bk,l}^{\phantom{*}}, 
\end{eqnarray}
where $n=\pm$ labels the $n-$th Bloch band and $\Delta \mathbf{x}_l = \mathbf{x}_l - \tilde{\mathbf{x}}_l$ is the relative shift between two embeddings.

Note that similar to Eq.~(12) of Ref.~\cite{Simon2020}, the generalized Berry connection is sensitive to the orbital embedding and transforms as 
\begin{equation}
    \mathbf{A}_{nm}(\bk) \rightarrow \mathbf{A}_{nm}(\bk)+\mathbf{\delta x}_{nm}(\bk)
\end{equation}
upon shifting the orbital positions by $\Delta \mathbf{x}_l$. For a two-band system, the quantum geometric tensor can then be written as 
\begin{equation}
    \mathcal{G}_{ij}(\bk)={A^i_{+-}(\bk)}^* A^j_{+-}(\bk)
\end{equation}
and thus transforms according to 
\begin{equation}
    \mathcal{G}_{ij}(\bk)\rightarrow 
    \mathcal{G}_{ij}(\bk)+ 
    {\delta x^i_{+-}(\bk)}^* A^j_{+-}(\bk) +
    {A^i_{+-}(\bk)}^* \delta x^j_{+-}(\bk) +
    {\delta x^i_{+-}(\bk)}^* \delta x^j_{+-}(\bk) .
\end{equation}
Note that for uniform shifts of the orbitals, i.e., $\Delta \mathbf{x}_l=\Delta \mathbf{x}$, we have $\mathbf{\delta x}_{+-}(\bk) = \sum_l \Delta \mathbf{x}_l u_{+,\bk,l}^* u_{-,\bk,l}^{\phantom{*}} = \Delta \mathbf{x} \, \langle u_{+,\bk} | u_{-,\bk}\rangle = 0$. Hence, this formula explicitly shows that the geometric tensor is left invariant; only relative shifts of the orbitals change the quantum metric. 

Let us now turn to the momentum averaged geometric tensor, where $\mathcal{G}^{(0)}_{ij}$ refers to the embedding with ${\bf x}_0={\bf0}$ and ${\bf x}_1=-a{\bf e}_z$ as in the main text and $\mathcal{G}^{(1)}_{ij}$ to an arbitrary embedding $\tilde{\mathbf{x}}_l$:
\begin{eqnarray}
    \frac{a^2}{(2\pi)^2}\iint_0^{2\pi/a}dk_x dk_y \mathcal{G}^{(1)}_{ij}(\bk,s) &=& \frac{a^2}{(2\pi)^2}\iint_0^{2\pi/a}dk_x dk_y \mathcal{G}^{(0)}_{ij}(\bk,s) + \frac{1}{4} (\tilde{x}_1-\tilde{x}_2)^i(\tilde{x}_1-\tilde{x}_2)^j \\
    &=& \frac{\zeta^2}{8} a^2 \delta_{ij} + \frac{1}{4} (\tilde{x}_1-\tilde{x}_2)^i(\tilde{x}_1-\tilde{x}_2)^j,
\end{eqnarray}
where we used $\mathbf{\delta x}_{+-}(\bk) = \frac{1}{2} (\Delta \mathbf{x}_1 - \Delta \mathbf{x}_2)$.
Hence, finite $x$- or $y$-components of $\tilde{\mathbf{x}}_l$ can only increase the diagonal elements, and thus the naive embedding as used in the manuscript minimizes the trace of the averaged metric and the associated Wannier orbitals are maximally localized.

\section{Absence of $O(\zeta)$ terms in projected Hamiltonian}
\label{app:linear-zeta}

Here we address the origin of the absence of terms in the projected Hamiltonian that are linear in $\zeta$. Those contributions vanish due to the chiral symmetry and the orbital independent interaction. For example, a density-assisted hopping of the form $\hat{d}^{\dagger}_{\r + {\bf x},s} \hat{d}^{\phantom{\dagger}}_{\r,s} \hat{d}^{\dagger}_{\r,s'} \hat{d}^{\phantom{\dagger}}_{\r,s'}$ is not generated as the contributions from both orbitals differ exactly by a sign and cancel each other. These perturbations remain a higher order correction in case of a weakly broken chiral symmetry or orbital-dependent interaction strength. The above operator generates odd parity sites and is therefore suppressed in $1/U$. In case of orbital dependent $U_l$, the relevant perturbation is of order $\zeta^2 (U_1-U_2)^2/(U_1+U_2)$.

\section{Correlation length of the charge density wave and supersolidity}
\label{app:sc-length}

\begin{figure}
	\begin{center}
		\includegraphics[width=0.75\columnwidth]{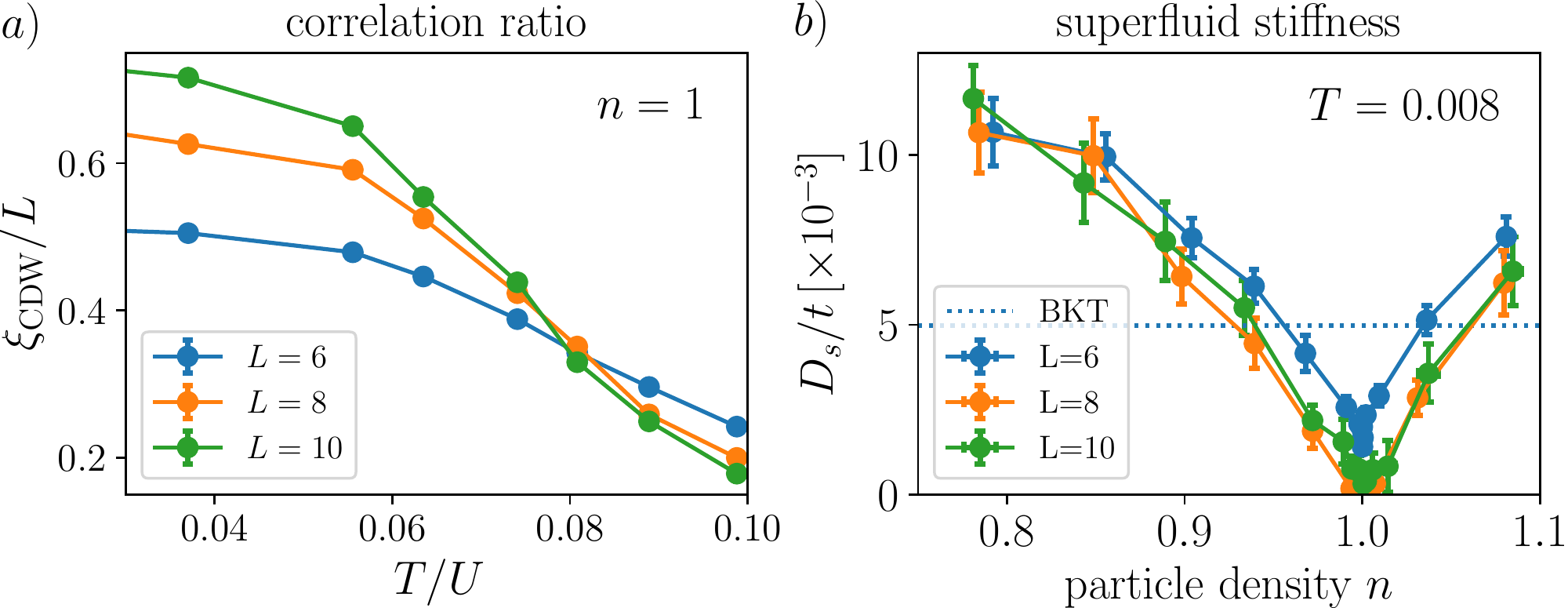} 
	\end{center}
	\caption{\label{fig:CDW} (a) CDW correlation length, $\xi_{\tn{CDW}}$, at quarter filling indicating a phase transition at $T_c=0.08\pm0.01$ for a fixed $V=0.08$ and $\zeta=0.75$. (b) Superfluid stiffness, $D_s$, at $T=0.008$ supporting a supersolid at finite carrier densities with $n_c=0.923\pm0.006$ and $n_c=1.062\pm0.008$ for hole and electron doping, respectively. 
	}
\end{figure}

We analyze the correlation length of the charge density wave, shown in Fig.~\ref{fig:CDW}(a). It is defined as 
\begin{equation}
	\xi_{\tn{CDW}}=\left(2\sin(\pi/L)\right)^{-1}\sqrt{\frac{S_N((\pi,\pi))}{S_N((\pi,\pi)+\delta\bq)}-1}\,,
\end{equation}
where $\delta\bq$ is a smallest non-vanishing momentum of the lattice and $S_N(\bq)$ is the equal-time density correlation function~\cite{Toldin15}. The ratio $\xi_{\tn{CDW}}/L$ increases (decreases) with system size in long-range ordered (disordered) phases. The ratio is an RG-invariant quantity such that the crossing point between different lattice sizes locates the phase transition.

At a low temperature, $T=0.008$, we calculate $D_s$ for a range of fillings near $n=1$ (Fig.~\ref{fig:CDW}b); for reference, we also draw the line $(2T/\pi)$, which determines the superconducting $T_c$. 
We find the critical carrier densities beyond which the ground state exhibits superconducting order to be $n_c=0.923\pm0.006$ and $n_c=1.062\pm0.008$ for hole and electron doping, respectively. We remark that the non-vanishing stiffness $D_s$ in the normal state is due to finite size effects~\cite{JHEBDC}.

\section{Single- and two-particle gap}

\begin{figure*}
	\begin{center}
          \includegraphics[width=\textwidth]{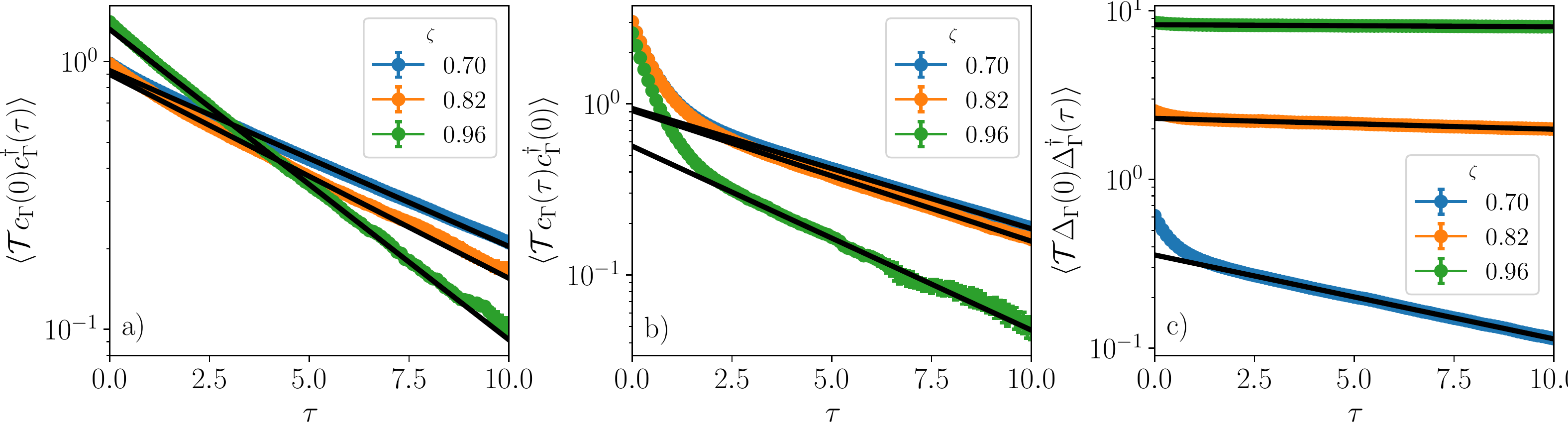}
	\end{center}
	\caption{\label{fig:gap-extraction} Imaginary-time displaced correlation functions for representatitive $\zeta$ and respective fits used to extract single- and two-particle excitation energies $\Delta_p$ and $\Delta_{pp}$: (a) $\langle \mathcal{T} c_{\Gamma }(0  ) c^{\dagger}_{\Gamma  }(\tau) \rangle $ for hole, (b) $\langle \mathcal{T} c_{\Gamma }(\tau  ) c^{\dagger}_{\Gamma  }(0) \rangle $ for electron, and (c) $\langle \mathcal{T} \Delta_{\Gamma }(0  ) \Delta^{\dagger}_{\Gamma  }(\tau) \rangle + \langle \mathcal{T} \Delta^{\dagger}_{\Gamma }(0  ) \Delta_{\Gamma  }(\tau) \rangle $ for two-particle excitations.
	}
\end{figure*}

We use the imaginary-time displaced correlation functions $\langle \mathcal{T} c_{\Gamma }(0  ) c^{\dagger}_{\Gamma  }(\tau) \rangle $ and $\langle \mathcal{T} c_{\Gamma }(\tau  ) c^{\dagger}_{\Gamma  }(0) \rangle $ to extract the single particle gap $\Delta_p$, and the symmetrized pair correlation function $\langle \mathcal{T} \Delta_{\Gamma }(0  ) \Delta^{\dagger}_{\Gamma  }(\tau) \rangle + \langle \mathcal{T} \Delta^{\dagger}_{\Gamma }(0  ) \Delta_{\Gamma  }(\tau) \rangle $ to extract the two-particle gap. In both cases, we focus on the center of the BZ, $\Gamma$, where the band minimum is located. We employ the ground state version of the DQMC algorithm where a trial wave function $|\Psi_{\mathrm{T}}\rangle$ is projected onto the ground state $|\Psi_{\mathrm{GS}}\rangle=e^{-\Theta \hat{H}}|\Psi_{\mathrm{T}}\rangle$ with $\Theta=30$. Hence, the correlation function decays exponentially with imaginary time $\tau$ where the decay rate is set by the excitation energies, e.g., $\langle \mathcal{T} c_{\Gamma }(\tau  ) c^{\dagger}_{\Gamma  }(0) \rangle = \sum_{\Psi_n} e^{-\tau (E_n-E_0)} |\langle \Psi_n | c^{\dagger}_{\Gamma  } |  \Psi_{\mathrm{GS}}\rangle|^2$, where $|\Psi_n\rangle$ is the $n^{\mathrm{th}}$ eigenstate of the interacting Hamiltonian $\hat{H}$ with energy $E_n$ and $E_0$ refers to the ground state energy. Note that this projective QMC algorithm is based on the canonical ensemble with fixed particle numbers. Here, we choose a trial wave function at quarter filling $n=1$ with $N=L^2/4$ particles. Since $c^\dagger_\Gamma$ creates an additional particle, $\Psi_n$ has to contain $N+1$ particles, and the decay rate in the long imaginary time limit determines $E_{N+1} - E_{N}$. Similarly, $\langle \mathcal{T} c_{\Gamma }(0  ) c^{\dagger}_{\Gamma  }(\tau) \rangle $ provides access to $E_{N-1} - E_{N}$. We extract the decay rate by fitting the exponentially decaying tail of the correlation function as shown in Fig.~\ref{fig:gap-extraction}(a),(b) and combine them to be independent of the chemical potential as $\Delta_p=0.5(E_{N+1} + E_{N-1} - 2 E_{N})$.

We use the symmetrized pair correlation function to extract the two-particle gap $\Delta_{pp}$ by fitting the exponentially decaying tail as shown in Fig.~\ref{fig:gap-extraction}(c). 
Strictly speaking, we should again determine the two correlation function $\langle \mathcal{T} \Delta_{\Gamma }(0  ) \Delta^{\dagger}_{\Gamma  }(\tau) \rangle$ and $\langle \mathcal{T} \Delta^{\dagger}_{\Gamma }(0  ) \Delta_{\Gamma  }(\tau) \rangle $ separately, determine their decay rates and combine both results to $\Delta_{pp}=0.5(E_{N+2} + E_{N-2} - 2 E_{N})$. Instead, we average both correlation function before extracting the decay rates. This is justified due to the similarity of both excitations energies and the corresponding exponential decay rate, directly visible in Fig.~\ref{fig:gap-extraction}(c), in particular for $\zeta=0.82$ and $\zeta=0.92$ where the two-particle excitation gap vanishes in the supersolid phase.

Finally, let us derive the analytical expression for the single-particle gap, $\Delta_p=\frac{U}{4} + 2V-\frac{7}{8}U\zeta^2$, in the small$-\zeta$ limit. When $\zeta=0$, all the terms in the projected Hamiltonian commute and the fully-polarized CDW states are the ground states. The additional electron can be added at the empty sites, and the energy cost is given by the Hubbard interaction, contributing $\frac{U_{\tn{eff}}}{2}$, and the energy of a defect with respect to the CDW, $-4\Delta^2_{\tn{CDW}}J_z$. For a finite $\zeta$, the quantum fluctuations reduce the order parameter $\Delta_{\tn{CDW}}=1/2-\mathcal{O}(\zeta^2)$. Additionally, the density-assisted hopping terms allow the defects to delocalize, and the dispersion is given by $-\frac{U \zeta^2 \Delta_{\tn{CDW}}}{4} \Big(2\sum_{a=\pm}\cos{k_a} + \sum_{a=x,y}\cos{2 k_a}\Big)$, where the density operator, $\eta^z_{\mathbf{i}}$, is replaced by its expectation value. Note that the band bottom is located at $\bk^{min}=0$ and $\bk^{min}=(\pi,\pi)$. Altogether, we have
\begin{eqnarray}
    \Delta_p &=& \frac{U_{\tn{eff}}}{2} -4\Delta^2_{\tn{CDW}}J_z -\frac{U \zeta^2 \Delta_{\tn{CDW}}}{4} \Big(2\sum_{a=\pm}\cos{k^{min}_a} + \sum_{a=x,y}\cos{2 k^{min}_a}\Big) \\
     &=& \frac{U}{4} - \frac{U\zeta^2}{8} - J_z -\frac{U \zeta^2 }{8} 6 \\
     &=& \frac{U}{4} + 2 V - \frac{7}{8}U \zeta^2 \, .
\end{eqnarray}

\end{widetext}

%% file: main.bbl
\begin{thebibliography}{43}%
\makeatletter
\providecommand \@ifxundefined [1]{%
 \@ifx{#1\undefined}
}%
\providecommand \@ifnum [1]{%
 \ifnum #1\expandafter \@firstoftwo
 \else \expandafter \@secondoftwo
 \fi
}%
\providecommand \@ifx [1]{%
 \ifx #1\expandafter \@firstoftwo
 \else \expandafter \@secondoftwo
 \fi
}%
\providecommand \natexlab [1]{#1}%
\providecommand \enquote  [1]{``#1''}%
\providecommand \bibnamefont  [1]{#1}%
\providecommand \bibfnamefont [1]{#1}%
\providecommand \citenamefont [1]{#1}%
\providecommand \href@noop [0]{\@secondoftwo}%
\providecommand \href [0]{\begingroup \@sanitize@url \@href}%
\providecommand \@href[1]{\@@startlink{#1}\@@href}%
\providecommand \@@href[1]{\endgroup#1\@@endlink}%
\providecommand \@sanitize@url [0]{\catcode `\\12\catcode `\$12\catcode
  `\&12\catcode `\#12\catcode `\^12\catcode `\_12\catcode `\%12\relax}%
\providecommand \@@startlink[1]{}%
\providecommand \@@endlink[0]{}%
\providecommand \url  [0]{\begingroup\@sanitize@url \@url }%
\providecommand \@url [1]{\endgroup\@href {#1}{\urlprefix }}%
\providecommand \urlprefix  [0]{URL }%
\providecommand \Eprint [0]{\href }%
\providecommand \doibase [0]{https://doi.org/}%
\providecommand \selectlanguage [0]{\@gobble}%
\providecommand \bibinfo  [0]{\@secondoftwo}%
\providecommand \bibfield  [0]{\@secondoftwo}%
\providecommand \translation [1]{[#1]}%
\providecommand \BibitemOpen [0]{}%
\providecommand \bibitemStop [0]{}%
\providecommand \bibitemNoStop [0]{.\EOS\space}%
\providecommand \EOS [0]{\spacefactor3000\relax}%
\providecommand \BibitemShut  [1]{\csname bibitem#1\endcsname}%
\let\auto@bib@innerbib\@empty
%</preamble>
\bibitem [{\citenamefont {Andrei}\ \emph {et~al.}(2021)\citenamefont {Andrei},
  \citenamefont {Efetov}, \citenamefont {Jarillo-Herrero}, \citenamefont
  {MacDonald}, \citenamefont {Mak}, \citenamefont {Senthil}, \citenamefont
  {Tutuc}, \citenamefont {Yazdani},\ and\ \citenamefont {Young}}]{moire_rev}%
  \BibitemOpen
  \bibfield  {author} {\bibinfo {author} {\bibfnamefont {E.~Y.}\ \bibnamefont
  {Andrei}}, \bibinfo {author} {\bibfnamefont {D.~K.}\ \bibnamefont {Efetov}},
  \bibinfo {author} {\bibfnamefont {P.}~\bibnamefont {Jarillo-Herrero}},
  \bibinfo {author} {\bibfnamefont {A.~H.}\ \bibnamefont {MacDonald}}, \bibinfo
  {author} {\bibfnamefont {K.~F.}\ \bibnamefont {Mak}}, \bibinfo {author}
  {\bibfnamefont {T.}~\bibnamefont {Senthil}}, \bibinfo {author} {\bibfnamefont
  {E.}~\bibnamefont {Tutuc}}, \bibinfo {author} {\bibfnamefont
  {A.}~\bibnamefont {Yazdani}},\ and\ \bibinfo {author} {\bibfnamefont {A.~F.}\
  \bibnamefont {Young}},\ }\bibfield  {title} {\bibinfo {title} {The marvels of
  moir{\'e} materials},\ }\href {https://doi.org/10.1038/s41578-021-00284-1}
  {\bibfield  {journal} {\bibinfo  {journal} {Nat. Rev. Mater.}\ }\textbf
  {\bibinfo {volume} {6}},\ \bibinfo {pages} {201} (\bibinfo {year}
  {2021})}\BibitemShut {NoStop}%
\bibitem [{\citenamefont {T{\"{o}}rm{\"{a}}}\ \emph {et~al.}(2022)\citenamefont
  {T{\"{o}}rm{\"{a}}}, \citenamefont {Peotta},\ and\ \citenamefont
  {Bernevig}}]{Bernevig_review}%
  \BibitemOpen
  \bibfield  {author} {\bibinfo {author} {\bibfnamefont {P.}~\bibnamefont
  {T{\"{o}}rm{\"{a}}}}, \bibinfo {author} {\bibfnamefont {S.}~\bibnamefont
  {Peotta}},\ and\ \bibinfo {author} {\bibfnamefont {B.~A.}\ \bibnamefont
  {Bernevig}},\ }\bibfield  {title} {\bibinfo {title} {{Superconductivity,
  superfluidity and quantum geometry in twisted multilayer systems}},\ }\href
  {https://doi.org/10.1038/s42254-022-00466-y} {\bibfield  {journal} {\bibinfo
  {journal} {Nat. Rev. Phys.}\ }\textbf {\bibinfo {volume} {4}},\ \bibinfo
  {pages} {528} (\bibinfo {year} {2022})},\ \Eprint
  {https://arxiv.org/abs/2111.00807} {arXiv:2111.00807} \BibitemShut {NoStop}%
\bibitem [{\citenamefont {Shaginyan}\ and\ \citenamefont
  {Khodel}(1990)}]{Shaginyan1990}%
  \BibitemOpen
  \bibfield  {author} {\bibinfo {author} {\bibfnamefont {V.}~\bibnamefont
  {Shaginyan}}\ and\ \bibinfo {author} {\bibfnamefont {V.}~\bibnamefont
  {Khodel}},\ }\bibfield  {title} {\bibinfo {title} {Superfluidity in system
  with fermion condensate},\ }\href
  {http://jetpletters.ac.ru/ps/1143/article_17312.shtml} {\bibfield  {journal}
  {\bibinfo  {journal} {JETP Lett}\ }\textbf {\bibinfo {volume} {51}} (\bibinfo
  {year} {1990})}\BibitemShut {NoStop}%
\bibitem [{\citenamefont {{Heikkil{\"a}}}\ \emph {et~al.}(2011)\citenamefont
  {{Heikkil{\"a}}}, \citenamefont {{Kopnin}},\ and\ \citenamefont
  {{Volovik}}}]{Volovik2011}%
  \BibitemOpen
  \bibfield  {author} {\bibinfo {author} {\bibfnamefont {T.~T.}\ \bibnamefont
  {{Heikkil{\"a}}}}, \bibinfo {author} {\bibfnamefont {N.~B.}\ \bibnamefont
  {{Kopnin}}},\ and\ \bibinfo {author} {\bibfnamefont {G.~E.}\ \bibnamefont
  {{Volovik}}},\ }\bibfield  {title} {\bibinfo {title} {{Flat bands in
  topological media}},\ }\href {https://doi.org/10.1134/S0021364011150045}
  {\bibfield  {journal} {\bibinfo  {journal} {Sov. J. Exp. Theor. Phys. Lett.}\
  }\textbf {\bibinfo {volume} {94}},\ \bibinfo {pages} {233} (\bibinfo {year}
  {2011})},\ \Eprint {https://arxiv.org/abs/1012.0905} {arXiv:1012.0905
  [cond-mat.str-el]} \BibitemShut {NoStop}%
\bibitem [{\citenamefont {Volovik}(2013)}]{Volovik2013}%
  \BibitemOpen
  \bibfield  {author} {\bibinfo {author} {\bibfnamefont {G.~E.}\ \bibnamefont
  {Volovik}},\ }\bibfield  {title} {\bibinfo {title} {Flat band in topological
  matter},\ }\href {https://doi.org/10.1007/s10948-013-2221-5} {\bibfield
  {journal} {\bibinfo  {journal} {J. Supercond. Nov. Magn.}\ }\textbf {\bibinfo
  {volume} {26}},\ \bibinfo {pages} {2887} (\bibinfo {year}
  {2013})}\BibitemShut {NoStop}%
\bibitem [{\citenamefont {Marzari}\ \emph {et~al.}(2012)\citenamefont
  {Marzari}, \citenamefont {Mostofi}, \citenamefont {Yates}, \citenamefont
  {Souza},\ and\ \citenamefont {Vanderbilt}}]{VanderbiltRMP}%
  \BibitemOpen
  \bibfield  {author} {\bibinfo {author} {\bibfnamefont {N.}~\bibnamefont
  {Marzari}}, \bibinfo {author} {\bibfnamefont {A.~A.}\ \bibnamefont
  {Mostofi}}, \bibinfo {author} {\bibfnamefont {J.~R.}\ \bibnamefont {Yates}},
  \bibinfo {author} {\bibfnamefont {I.}~\bibnamefont {Souza}},\ and\ \bibinfo
  {author} {\bibfnamefont {D.}~\bibnamefont {Vanderbilt}},\ }\bibfield  {title}
  {\bibinfo {title} {Maximally localized wannier functions: Theory and
  applications},\ }\href {https://doi.org/10.1103/RevModPhys.84.1419}
  {\bibfield  {journal} {\bibinfo  {journal} {Rev. Mod. Phys.}\ }\textbf
  {\bibinfo {volume} {84}},\ \bibinfo {pages} {1419} (\bibinfo {year}
  {2012})}\BibitemShut {NoStop}%
\bibitem [{\citenamefont {Peotta}\ and\ \citenamefont
  {T{\"o}rm{\"a}}(2015)}]{Torma15}%
  \BibitemOpen
  \bibfield  {author} {\bibinfo {author} {\bibfnamefont {S.}~\bibnamefont
  {Peotta}}\ and\ \bibinfo {author} {\bibfnamefont {P.}~\bibnamefont
  {T{\"o}rm{\"a}}},\ }\bibfield  {title} {\bibinfo {title} {Superfluidity in
  topologically nontrivial flat bands},\ }\href
  {https://doi.org/10.1038/ncomms9944} {\bibfield  {journal} {\bibinfo
  {journal} {Nat. Commun.}\ }\textbf {\bibinfo {volume} {6}},\ \bibinfo {pages}
  {8944} (\bibinfo {year} {2015})}\BibitemShut {NoStop}%
\bibitem [{\citenamefont {Blankenbecler}\ \emph {et~al.}(1981)\citenamefont
  {Blankenbecler}, \citenamefont {Scalapino},\ and\ \citenamefont
  {Sugar}}]{Blankenbecler1981}%
  \BibitemOpen
  \bibfield  {author} {\bibinfo {author} {\bibfnamefont {R.}~\bibnamefont
  {Blankenbecler}}, \bibinfo {author} {\bibfnamefont {D.~J.}\ \bibnamefont
  {Scalapino}},\ and\ \bibinfo {author} {\bibfnamefont {R.~L.}\ \bibnamefont
  {Sugar}},\ }\bibfield  {title} {\bibinfo {title} {Monte carlo calculations of
  coupled boson-fermion systems. i},\ }\href
  {https://doi.org/10.1103/PhysRevD.24.2278} {\bibfield  {journal} {\bibinfo
  {journal} {Phys. Rev. D}\ }\textbf {\bibinfo {volume} {24}},\ \bibinfo
  {pages} {2278} (\bibinfo {year} {1981})}\BibitemShut {NoStop}%
\bibitem [{\citenamefont {Hofmann}\ \emph {et~al.}(2020)\citenamefont
  {Hofmann}, \citenamefont {Berg},\ and\ \citenamefont {Chowdhury}}]{JHEBDC}%
  \BibitemOpen
  \bibfield  {author} {\bibinfo {author} {\bibfnamefont {J.~S.}\ \bibnamefont
  {Hofmann}}, \bibinfo {author} {\bibfnamefont {E.}~\bibnamefont {Berg}},\ and\
  \bibinfo {author} {\bibfnamefont {D.}~\bibnamefont {Chowdhury}},\ }\bibfield
  {title} {\bibinfo {title} {Superconductivity, pseudogap, and phase separation
  in topological flat bands},\ }\href
  {https://doi.org/10.1103/PhysRevB.102.201112} {\bibfield  {journal} {\bibinfo
   {journal} {Phys. Rev. B}\ }\textbf {\bibinfo {volume} {102}},\ \bibinfo
  {pages} {201112} (\bibinfo {year} {2020})}\BibitemShut {NoStop}%
\bibitem [{\citenamefont {Peri}\ \emph {et~al.}(2021)\citenamefont {Peri},
  \citenamefont {Song}, \citenamefont {Bernevig},\ and\ \citenamefont
  {Huber}}]{Huber21}%
  \BibitemOpen
  \bibfield  {author} {\bibinfo {author} {\bibfnamefont {V.}~\bibnamefont
  {Peri}}, \bibinfo {author} {\bibfnamefont {Z.-D.}\ \bibnamefont {Song}},
  \bibinfo {author} {\bibfnamefont {B.~A.}\ \bibnamefont {Bernevig}},\ and\
  \bibinfo {author} {\bibfnamefont {S.~D.}\ \bibnamefont {Huber}},\ }\bibfield
  {title} {\bibinfo {title} {Fragile topology and flat-band superconductivity
  in the strong-coupling regime},\ }\href
  {https://doi.org/10.1103/PhysRevLett.126.027002} {\bibfield  {journal}
  {\bibinfo  {journal} {Phys. Rev. Lett.}\ }\textbf {\bibinfo {volume} {126}},\
  \bibinfo {pages} {027002} (\bibinfo {year} {2021})}\BibitemShut {NoStop}%
\bibitem [{\citenamefont {Zhang}\ \emph {et~al.}(2022)\citenamefont {Zhang},
  \citenamefont {Sun}, \citenamefont {Li}, \citenamefont {Pan},\ and\
  \citenamefont {Meng}}]{Zhang2021}%
  \BibitemOpen
  \bibfield  {author} {\bibinfo {author} {\bibfnamefont {X.}~\bibnamefont
  {Zhang}}, \bibinfo {author} {\bibfnamefont {K.}~\bibnamefont {Sun}}, \bibinfo
  {author} {\bibfnamefont {H.}~\bibnamefont {Li}}, \bibinfo {author}
  {\bibfnamefont {G.}~\bibnamefont {Pan}},\ and\ \bibinfo {author}
  {\bibfnamefont {Z.~Y.}\ \bibnamefont {Meng}},\ }\bibfield  {title} {\bibinfo
  {title} {{Superconductivity and bosonic fluid emerging from moir{\'{e}} flat
  bands}},\ }\href {https://doi.org/10.1103/PhysRevB.106.184517} {\bibfield
  {journal} {\bibinfo  {journal} {Phys. Rev. B}\ }\textbf {\bibinfo {volume}
  {106}},\ \bibinfo {pages} {184517} (\bibinfo {year} {2022})},\ \Eprint
  {https://arxiv.org/abs/2111.10018v3} {arXiv:2111.10018v3} \BibitemShut
  {NoStop}%
\bibitem [{\citenamefont {Herzog-Arbeitman}\ \emph {et~al.}(2022)\citenamefont
  {Herzog-Arbeitman}, \citenamefont {Peri}, \citenamefont {Schindler},
  \citenamefont {Huber},\ and\ \citenamefont {Bernevig}}]{Bernevig21}%
  \BibitemOpen
  \bibfield  {author} {\bibinfo {author} {\bibfnamefont {J.}~\bibnamefont
  {Herzog-Arbeitman}}, \bibinfo {author} {\bibfnamefont {V.}~\bibnamefont
  {Peri}}, \bibinfo {author} {\bibfnamefont {F.}~\bibnamefont {Schindler}},
  \bibinfo {author} {\bibfnamefont {S.~D.}\ \bibnamefont {Huber}},\ and\
  \bibinfo {author} {\bibfnamefont {B.~A.}\ \bibnamefont {Bernevig}},\
  }\bibfield  {title} {\bibinfo {title} {{Superfluid Weight Bounds from
  Symmetry and Quantum Geometry in Flat Bands}},\ }\href
  {https://doi.org/10.1103/PhysRevLett.128.087002} {\bibfield  {journal}
  {\bibinfo  {journal} {Phys. Rev. Lett.}\ }\textbf {\bibinfo {volume} {128}},\
  \bibinfo {pages} {087002} (\bibinfo {year} {2022})},\ \Eprint
  {https://arxiv.org/abs/2110.14663} {arXiv:2110.14663} \BibitemShut {NoStop}%
\bibitem [{\citenamefont {Tovmasyan}\ \emph {et~al.}(2016)\citenamefont
  {Tovmasyan}, \citenamefont {Peotta}, \citenamefont {T\"orm\"a},\ and\
  \citenamefont {Huber}}]{Torma16b}%
  \BibitemOpen
  \bibfield  {author} {\bibinfo {author} {\bibfnamefont {M.}~\bibnamefont
  {Tovmasyan}}, \bibinfo {author} {\bibfnamefont {S.}~\bibnamefont {Peotta}},
  \bibinfo {author} {\bibfnamefont {P.}~\bibnamefont {T\"orm\"a}},\ and\
  \bibinfo {author} {\bibfnamefont {S.~D.}\ \bibnamefont {Huber}},\ }\bibfield
  {title} {\bibinfo {title} {Effective theory and emergent $\text{SU}(2)$
  symmetry in the flat bands of attractive hubbard models},\ }\href
  {https://doi.org/10.1103/PhysRevB.94.245149} {\bibfield  {journal} {\bibinfo
  {journal} {Phys. Rev. B}\ }\textbf {\bibinfo {volume} {94}},\ \bibinfo
  {pages} {245149} (\bibinfo {year} {2016})}\BibitemShut {NoStop}%
\bibitem [{\citenamefont {Xie}\ \emph {et~al.}(2020)\citenamefont {Xie},
  \citenamefont {Song}, \citenamefont {Lian},\ and\ \citenamefont
  {Bernevig}}]{Bernevig19}%
  \BibitemOpen
  \bibfield  {author} {\bibinfo {author} {\bibfnamefont {F.}~\bibnamefont
  {Xie}}, \bibinfo {author} {\bibfnamefont {Z.}~\bibnamefont {Song}}, \bibinfo
  {author} {\bibfnamefont {B.}~\bibnamefont {Lian}},\ and\ \bibinfo {author}
  {\bibfnamefont {B.~A.}\ \bibnamefont {Bernevig}},\ }\bibfield  {title}
  {\bibinfo {title} {Topology-bounded superfluid weight in twisted bilayer
  graphene},\ }\href {https://doi.org/10.1103/PhysRevLett.124.167002}
  {\bibfield  {journal} {\bibinfo  {journal} {Phys. Rev. Lett.}\ }\textbf
  {\bibinfo {volume} {124}},\ \bibinfo {pages} {167002} (\bibinfo {year}
  {2020})}\BibitemShut {NoStop}%
\bibitem [{\citenamefont {Hu}\ \emph {et~al.}(2019)\citenamefont {Hu},
  \citenamefont {Hyart}, \citenamefont {Pikulin},\ and\ \citenamefont
  {Rossi}}]{Rossi19}%
  \BibitemOpen
  \bibfield  {author} {\bibinfo {author} {\bibfnamefont {X.}~\bibnamefont
  {Hu}}, \bibinfo {author} {\bibfnamefont {T.}~\bibnamefont {Hyart}}, \bibinfo
  {author} {\bibfnamefont {D.~I.}\ \bibnamefont {Pikulin}},\ and\ \bibinfo
  {author} {\bibfnamefont {E.}~\bibnamefont {Rossi}},\ }\bibfield  {title}
  {\bibinfo {title} {Geometric and conventional contribution to the superfluid
  weight in twisted bilayer graphene},\ }\href
  {https://doi.org/10.1103/PhysRevLett.123.237002} {\bibfield  {journal}
  {\bibinfo  {journal} {Phys. Rev. Lett.}\ }\textbf {\bibinfo {volume} {123}},\
  \bibinfo {pages} {237002} (\bibinfo {year} {2019})}\BibitemShut {NoStop}%
\bibitem [{\citenamefont {Julku}\ \emph {et~al.}(2020)\citenamefont {Julku},
  \citenamefont {Peltonen}, \citenamefont {Liang}, \citenamefont {Heikkil\"a},\
  and\ \citenamefont {T\"orm\"a}}]{Torma19}%
  \BibitemOpen
  \bibfield  {author} {\bibinfo {author} {\bibfnamefont {A.}~\bibnamefont
  {Julku}}, \bibinfo {author} {\bibfnamefont {T.~J.}\ \bibnamefont {Peltonen}},
  \bibinfo {author} {\bibfnamefont {L.}~\bibnamefont {Liang}}, \bibinfo
  {author} {\bibfnamefont {T.~T.}\ \bibnamefont {Heikkil\"a}},\ and\ \bibinfo
  {author} {\bibfnamefont {P.}~\bibnamefont {T\"orm\"a}},\ }\bibfield  {title}
  {\bibinfo {title} {Superfluid weight and berezinskii-kosterlitz-thouless
  transition temperature of twisted bilayer graphene},\ }\href
  {https://doi.org/10.1103/PhysRevB.101.060505} {\bibfield  {journal} {\bibinfo
   {journal} {Phys. Rev. B}\ }\textbf {\bibinfo {volume} {101}},\ \bibinfo
  {pages} {060505} (\bibinfo {year} {2020})}\BibitemShut {NoStop}%
\bibitem [{\citenamefont {Hazra}\ \emph {et~al.}(2019)\citenamefont {Hazra},
  \citenamefont {Verma},\ and\ \citenamefont {Randeria}}]{Hazra18}%
  \BibitemOpen
  \bibfield  {author} {\bibinfo {author} {\bibfnamefont {T.}~\bibnamefont
  {Hazra}}, \bibinfo {author} {\bibfnamefont {N.}~\bibnamefont {Verma}},\ and\
  \bibinfo {author} {\bibfnamefont {M.}~\bibnamefont {Randeria}},\ }\bibfield
  {title} {\bibinfo {title} {Bounds on the superconducting transition
  temperature: Applications to twisted bilayer graphene and cold atoms},\
  }\href {https://doi.org/10.1103/PhysRevX.9.031049} {\bibfield  {journal}
  {\bibinfo  {journal} {Phys. Rev. X}\ }\textbf {\bibinfo {volume} {9}},\
  \bibinfo {pages} {031049} (\bibinfo {year} {2019})}\BibitemShut {NoStop}%
\bibitem [{\citenamefont {{Verma}}\ \emph {et~al.}(2021)\citenamefont
  {{Verma}}, \citenamefont {{Hazra}},\ and\ \citenamefont {{Randeria}}}]{MR21}%
  \BibitemOpen
  \bibfield  {author} {\bibinfo {author} {\bibfnamefont {N.}~\bibnamefont
  {{Verma}}}, \bibinfo {author} {\bibfnamefont {T.}~\bibnamefont {{Hazra}}},\
  and\ \bibinfo {author} {\bibfnamefont {M.}~\bibnamefont {{Randeria}}},\
  }\bibfield  {title} {\bibinfo {title} {{Optical spectral weight, phase
  stiffness, and $T_c$ bounds for trivial and topological flat band
  superconductors}},\ }\href {https://doi.org/10.1073/pnas.2106744118}
  {\bibfield  {journal} {\bibinfo  {journal} {Proc. Natl. Acad. Sci.}\ }\textbf
  {\bibinfo {volume} {118}},\ \bibinfo {pages} {e2106744118} (\bibinfo {year}
  {2021})}\BibitemShut {NoStop}%
\bibitem [{\citenamefont {Mao}\ and\ \citenamefont {Chowdhury}(2023)}]{DM21}%
  \BibitemOpen
  \bibfield  {author} {\bibinfo {author} {\bibfnamefont {D.}~\bibnamefont
  {Mao}}\ and\ \bibinfo {author} {\bibfnamefont {D.}~\bibnamefont
  {Chowdhury}},\ }\bibfield  {title} {\bibinfo {title} {Diamagnetic response
  and phase stiffness for interacting isolated narrow bands},\ }\href
  {https://doi.org/10.1073/pnas.2217816120} {\bibfield  {journal} {\bibinfo
  {journal} {Proc. Natl. Acad. Sci.}\ }\textbf {\bibinfo {volume} {120}},\
  \bibinfo {pages} {e2217816120} (\bibinfo {year} {2023})}\BibitemShut
  {NoStop}%
\bibitem [{\citenamefont {Hofmann}\ \emph {et~al.}(2022)\citenamefont
  {Hofmann}, \citenamefont {Chowdhury}, \citenamefont {Kivelson},\ and\
  \citenamefont {Berg}}]{boundless}%
  \BibitemOpen
  \bibfield  {author} {\bibinfo {author} {\bibfnamefont {J.~S.}\ \bibnamefont
  {Hofmann}}, \bibinfo {author} {\bibfnamefont {D.}~\bibnamefont {Chowdhury}},
  \bibinfo {author} {\bibfnamefont {S.~A.}\ \bibnamefont {Kivelson}},\ and\
  \bibinfo {author} {\bibfnamefont {E.}~\bibnamefont {Berg}},\ }\bibfield
  {title} {\bibinfo {title} {Heuristic bounds on superconductivity and how to
  exceed them},\ }\href {https://doi.org/10.1038/s41535-022-00491-1} {\bibfield
   {journal} {\bibinfo  {journal} {npj Quantum Mater.}\ }\textbf {\bibinfo
  {volume} {7}},\ \bibinfo {pages} {83} (\bibinfo {year} {2022})}\BibitemShut
  {NoStop}%
\bibitem [{si()}]{si}%
  \BibitemOpen
  \href@noop {} {\bibinfo {title} {See supplementary information for additional
  details on computations for the wannier wave function, the quantum geometric
  tensor and its connection with the orbital embedding, the projected
  hamiltonian, the correlation length for cdw, and the particle-excitation
  gaps.}}\BibitemShut {Stop}%
\bibitem [{\citenamefont {Simon}\ and\ \citenamefont
  {Rudner}(2020)}]{Simon2020}%
  \BibitemOpen
  \bibfield  {author} {\bibinfo {author} {\bibfnamefont {S.~H.}\ \bibnamefont
  {Simon}}\ and\ \bibinfo {author} {\bibfnamefont {M.~S.}\ \bibnamefont
  {Rudner}},\ }\bibfield  {title} {\bibinfo {title} {{Contrasting lattice
  geometry dependent versus independent quantities: Ramifications for Berry
  curvature, energy gaps, and dynamics}},\ }\href
  {https://doi.org/10.1103/PhysRevB.102.165148} {\bibfield  {journal} {\bibinfo
   {journal} {Phys. Rev. B}\ }\textbf {\bibinfo {volume} {102}},\ \bibinfo
  {pages} {165148} (\bibinfo {year} {2020})}\BibitemShut {NoStop}%
\bibitem [{\citenamefont {Bercx}\ \emph {et~al.}(2017)\citenamefont {Bercx},
  \citenamefont {Goth}, \citenamefont {Hofmann},\ and\ \citenamefont
  {Assaad}}]{ALF2017}%
  \BibitemOpen
  \bibfield  {author} {\bibinfo {author} {\bibfnamefont {M.}~\bibnamefont
  {Bercx}}, \bibinfo {author} {\bibfnamefont {F.}~\bibnamefont {Goth}},
  \bibinfo {author} {\bibfnamefont {J.~S.}\ \bibnamefont {Hofmann}},\ and\
  \bibinfo {author} {\bibfnamefont {F.}~\bibnamefont {Assaad}},\ }\bibfield
  {title} {\bibinfo {title} {The alf (algorithms for lattice fermions) project
  release 1.0. documentation for the auxiliary field quantum monte carlo
  code},\ }\href {https://doi.org/10.21468/SciPostPhys.3.2.013} {\bibfield
  {journal} {\bibinfo  {journal} {SciPost Phys.}\ }\textbf {\bibinfo {volume}
  {3}},\ \bibinfo {eid} {013} (\bibinfo {year} {2017})},\ \Eprint
  {https://arxiv.org/abs/1704.00131} {arXiv:1704.00131 [cond-mat.str-el]}
  \BibitemShut {NoStop}%
\bibitem [{\citenamefont {Assaad}\ \emph {et~al.}(2022)\citenamefont {Assaad},
  \citenamefont {Bercx}, \citenamefont {Goth}, \citenamefont {G{\"{o}}tz},
  \citenamefont {Hofmann}, \citenamefont {Huffman}, \citenamefont {Liu},
  \citenamefont {{Parisen Toldin}}, \citenamefont {Portela},\ and\
  \citenamefont {Schwab}}]{ALF2021}%
  \BibitemOpen
  \bibfield  {author} {\bibinfo {author} {\bibfnamefont {F.}~\bibnamefont
  {Assaad}}, \bibinfo {author} {\bibfnamefont {M.}~\bibnamefont {Bercx}},
  \bibinfo {author} {\bibfnamefont {F.}~\bibnamefont {Goth}}, \bibinfo {author}
  {\bibfnamefont {A.}~\bibnamefont {G{\"{o}}tz}}, \bibinfo {author}
  {\bibfnamefont {J.}~\bibnamefont {Hofmann}}, \bibinfo {author} {\bibfnamefont
  {E.}~\bibnamefont {Huffman}}, \bibinfo {author} {\bibfnamefont
  {Z.}~\bibnamefont {Liu}}, \bibinfo {author} {\bibfnamefont {F.}~\bibnamefont
  {{Parisen Toldin}}}, \bibinfo {author} {\bibfnamefont {J.}~\bibnamefont
  {Portela}},\ and\ \bibinfo {author} {\bibfnamefont {J.}~\bibnamefont
  {Schwab}},\ }\bibfield  {title} {\bibinfo {title} {{The ALF (Algorithms for
  Lattice Fermions) project release 2.0. Documentation for the auxiliary-field
  quantum Monte Carlo code}},\ }\href
  {https://doi.org/10.21468/SciPostPhysCodeb.1} {\bibfield  {journal} {\bibinfo
   {journal} {SciPost Phys. Codebases}\ ,\ \bibinfo {pages} {1}} (\bibinfo
  {year} {2022})},\ \Eprint {https://arxiv.org/abs/2012.11914}
  {arXiv:2012.11914} \BibitemShut {NoStop}%
\bibitem [{\citenamefont {Nelson}\ and\ \citenamefont
  {Kosterlitz}(1977)}]{NK77}%
  \BibitemOpen
  \bibfield  {author} {\bibinfo {author} {\bibfnamefont {D.~R.}\ \bibnamefont
  {Nelson}}\ and\ \bibinfo {author} {\bibfnamefont {J.~M.}\ \bibnamefont
  {Kosterlitz}},\ }\bibfield  {title} {\bibinfo {title} {Universal jump in the
  superfluid density of two-dimensional superfluids},\ }\href
  {https://doi.org/10.1103/PhysRevLett.39.1201} {\bibfield  {journal} {\bibinfo
   {journal} {Phys. Rev. Lett.}\ }\textbf {\bibinfo {volume} {39}},\ \bibinfo
  {pages} {1201} (\bibinfo {year} {1977})}\BibitemShut {NoStop}%
\bibitem [{\citenamefont {Scalapino}\ \emph {et~al.}(1993)\citenamefont
  {Scalapino}, \citenamefont {White},\ and\ \citenamefont
  {Zhang}}]{SFcriteria}%
  \BibitemOpen
  \bibfield  {author} {\bibinfo {author} {\bibfnamefont {D.~J.}\ \bibnamefont
  {Scalapino}}, \bibinfo {author} {\bibfnamefont {S.~R.}\ \bibnamefont
  {White}},\ and\ \bibinfo {author} {\bibfnamefont {S.}~\bibnamefont {Zhang}},\
  }\bibfield  {title} {\bibinfo {title} {Insulator, metal, or superconductor:
  The criteria},\ }\href {https://doi.org/10.1103/PhysRevB.47.7995} {\bibfield
  {journal} {\bibinfo  {journal} {Phys. Rev. B}\ }\textbf {\bibinfo {volume}
  {47}},\ \bibinfo {pages} {7995} (\bibinfo {year} {1993})}\BibitemShut
  {NoStop}%
\bibitem [{\citenamefont {Parisen~Toldin}\ \emph {et~al.}(2015)\citenamefont
  {Parisen~Toldin}, \citenamefont {Hohenadler}, \citenamefont {Assaad},\ and\
  \citenamefont {Herbut}}]{Toldin15}%
  \BibitemOpen
  \bibfield  {author} {\bibinfo {author} {\bibfnamefont {F.}~\bibnamefont
  {Parisen~Toldin}}, \bibinfo {author} {\bibfnamefont {M.}~\bibnamefont
  {Hohenadler}}, \bibinfo {author} {\bibfnamefont {F.~F.}\ \bibnamefont
  {Assaad}},\ and\ \bibinfo {author} {\bibfnamefont {I.~F.}\ \bibnamefont
  {Herbut}},\ }\bibfield  {title} {\bibinfo {title} {Fermionic quantum
  criticality in honeycomb and $\ensuremath{\pi}$-flux hubbard models:
  Finite-size scaling of renormalization-group-invariant observables from
  quantum monte carlo},\ }\href {https://doi.org/10.1103/PhysRevB.91.165108}
  {\bibfield  {journal} {\bibinfo  {journal} {Phys. Rev. B}\ }\textbf {\bibinfo
  {volume} {91}},\ \bibinfo {pages} {165108} (\bibinfo {year}
  {2015})}\BibitemShut {NoStop}%
\bibitem [{\citenamefont {Popov}(1972)}]{popov1972theory}%
  \BibitemOpen
  \bibfield  {author} {\bibinfo {author} {\bibfnamefont {V.~N.}\ \bibnamefont
  {Popov}},\ }\bibfield  {title} {\bibinfo {title} {{On the theory of the
  superfluidity of two- and one-dimensional bose systems}},\ }\href
  {https://doi.org/10.1007/BF01028373} {\bibfield  {journal} {\bibinfo
  {journal} {Theor. Math. Phys.}\ }\textbf {\bibinfo {volume} {11}},\ \bibinfo
  {pages} {565} (\bibinfo {year} {1972})}\BibitemShut {NoStop}%
\bibitem [{\citenamefont {Kagan}\ \emph {et~al.}(1987)\citenamefont {Kagan},
  \citenamefont {Svistunov},\ and\ \citenamefont
  {Shlyapnikov}}]{kagan1987influence}%
  \BibitemOpen
  \bibfield  {author} {\bibinfo {author} {\bibfnamefont {Y.}~\bibnamefont
  {Kagan}}, \bibinfo {author} {\bibfnamefont {B.}~\bibnamefont {Svistunov}},\
  and\ \bibinfo {author} {\bibfnamefont {G.}~\bibnamefont {Shlyapnikov}},\
  }\bibfield  {title} {\bibinfo {title} {Influence on inelastic processes of
  the phase transition in a weakly collisional two-dimensional bose gas},\
  }\href@noop {} {\bibfield  {journal} {\bibinfo  {journal} {Sov. Phys. JETP}\
  }\textbf {\bibinfo {volume} {66}},\ \bibinfo {pages} {314} (\bibinfo {year}
  {1987})}\BibitemShut {NoStop}%
\bibitem [{\citenamefont {Fisher}\ and\ \citenamefont
  {Hohenberg}(1988)}]{Fisher1988}%
  \BibitemOpen
  \bibfield  {author} {\bibinfo {author} {\bibfnamefont {D.~S.}\ \bibnamefont
  {Fisher}}\ and\ \bibinfo {author} {\bibfnamefont {P.~C.}\ \bibnamefont
  {Hohenberg}},\ }\bibfield  {title} {\bibinfo {title} {Dilute bose gas in two
  dimensions},\ }\href {https://doi.org/10.1103/PhysRevB.37.4936} {\bibfield
  {journal} {\bibinfo  {journal} {Phys. Rev. B}\ }\textbf {\bibinfo {volume}
  {37}},\ \bibinfo {pages} {4936} (\bibinfo {year} {1988})}\BibitemShut
  {NoStop}%
\bibitem [{\citenamefont {{Beach}}(2004)}]{MaxEnt2004}%
  \BibitemOpen
  \bibfield  {author} {\bibinfo {author} {\bibfnamefont {K.~S.~D.}\
  \bibnamefont {{Beach}}},\ }\bibfield  {title} {\bibinfo {title} {{Identifying
  the maximum entropy method as a special limit of stochastic analytic
  continuation}},\ }\href@noop {} {\bibfield  {journal} {\bibinfo  {journal}
  {arXiv e-prints}\ ,\ \bibinfo {eid} {cond-mat/0403055}} (\bibinfo {year}
  {2004})},\ \Eprint {https://arxiv.org/abs/cond-mat/0403055}
  {arXiv:cond-mat/0403055 [cond-mat.str-el]} \BibitemShut {NoStop}%
\bibitem [{\citenamefont {Julku}\ \emph {et~al.}(2016)\citenamefont {Julku},
  \citenamefont {Peotta}, \citenamefont {Vanhala}, \citenamefont {Kim},\ and\
  \citenamefont {T\"orm\"a}}]{Torma16a}%
  \BibitemOpen
  \bibfield  {author} {\bibinfo {author} {\bibfnamefont {A.}~\bibnamefont
  {Julku}}, \bibinfo {author} {\bibfnamefont {S.}~\bibnamefont {Peotta}},
  \bibinfo {author} {\bibfnamefont {T.~I.}\ \bibnamefont {Vanhala}}, \bibinfo
  {author} {\bibfnamefont {D.-H.}\ \bibnamefont {Kim}},\ and\ \bibinfo {author}
  {\bibfnamefont {P.}~\bibnamefont {T\"orm\"a}},\ }\bibfield  {title} {\bibinfo
  {title} {Geometric origin of superfluidity in the lieb-lattice flat band},\
  }\href {https://doi.org/10.1103/PhysRevLett.117.045303} {\bibfield  {journal}
  {\bibinfo  {journal} {Phys. Rev. Lett.}\ }\textbf {\bibinfo {volume} {117}},\
  \bibinfo {pages} {045303} (\bibinfo {year} {2016})}\BibitemShut {NoStop}%
\bibitem [{\citenamefont {Li}\ \emph {et~al.}(2017)\citenamefont {Li},
  \citenamefont {Lee}, \citenamefont {Huang}, \citenamefont {Burchesky},
  \citenamefont {Shteynas}, \citenamefont {Top}, \citenamefont {Jamison},\ and\
  \citenamefont {Ketterle}}]{Li17_supersolid_exp}%
  \BibitemOpen
  \bibfield  {author} {\bibinfo {author} {\bibfnamefont {J.-R.}\ \bibnamefont
  {Li}}, \bibinfo {author} {\bibfnamefont {J.}~\bibnamefont {Lee}}, \bibinfo
  {author} {\bibfnamefont {W.}~\bibnamefont {Huang}}, \bibinfo {author}
  {\bibfnamefont {S.}~\bibnamefont {Burchesky}}, \bibinfo {author}
  {\bibfnamefont {B.}~\bibnamefont {Shteynas}}, \bibinfo {author}
  {\bibfnamefont {F.~{\c{C}}.}\ \bibnamefont {Top}}, \bibinfo {author}
  {\bibfnamefont {A.~O.}\ \bibnamefont {Jamison}},\ and\ \bibinfo {author}
  {\bibfnamefont {W.}~\bibnamefont {Ketterle}},\ }\bibfield  {title} {\bibinfo
  {title} {{A stripe phase with supersolid properties in spin–orbit-coupled
  Bose–Einstein condensates}},\ }\href {https://doi.org/10.1038/nature21431}
  {\bibfield  {journal} {\bibinfo  {journal} {Nature}\ }\textbf {\bibinfo
  {volume} {543}},\ \bibinfo {pages} {91} (\bibinfo {year} {2017})}\BibitemShut
  {NoStop}%
\bibitem [{\citenamefont {L{\'{e}}onard}\ \emph {et~al.}(2017)\citenamefont
  {L{\'{e}}onard}, \citenamefont {Morales}, \citenamefont {Zupancic},
  \citenamefont {Esslinger},\ and\ \citenamefont
  {Donner}}]{Leonard17_supersolid_exp}%
  \BibitemOpen
  \bibfield  {author} {\bibinfo {author} {\bibfnamefont {J.}~\bibnamefont
  {L{\'{e}}onard}}, \bibinfo {author} {\bibfnamefont {A.}~\bibnamefont
  {Morales}}, \bibinfo {author} {\bibfnamefont {P.}~\bibnamefont {Zupancic}},
  \bibinfo {author} {\bibfnamefont {T.}~\bibnamefont {Esslinger}},\ and\
  \bibinfo {author} {\bibfnamefont {T.}~\bibnamefont {Donner}},\ }\bibfield
  {title} {\bibinfo {title} {{Supersolid formation in a quantum gas breaking a
  continuous translational symmetry}},\ }\href
  {https://doi.org/10.1038/nature21067} {\bibfield  {journal} {\bibinfo
  {journal} {Nature}\ }\textbf {\bibinfo {volume} {543}},\ \bibinfo {pages}
  {87} (\bibinfo {year} {2017})},\ \Eprint {https://arxiv.org/abs/1609.09053}
  {arXiv:1609.09053} \BibitemShut {NoStop}%
\bibitem [{\citenamefont {Tanzi}\ \emph {et~al.}(2019)\citenamefont {Tanzi},
  \citenamefont {Roccuzzo}, \citenamefont {Lucioni}, \citenamefont
  {Fam{\`{a}}}, \citenamefont {Fioretti}, \citenamefont {Gabbanini},
  \citenamefont {Modugno}, \citenamefont {Recati},\ and\ \citenamefont
  {Stringari}}]{Tanzi19_supersolid_exp}%
  \BibitemOpen
  \bibfield  {author} {\bibinfo {author} {\bibfnamefont {L.}~\bibnamefont
  {Tanzi}}, \bibinfo {author} {\bibfnamefont {S.~M.}\ \bibnamefont {Roccuzzo}},
  \bibinfo {author} {\bibfnamefont {E.}~\bibnamefont {Lucioni}}, \bibinfo
  {author} {\bibfnamefont {F.}~\bibnamefont {Fam{\`{a}}}}, \bibinfo {author}
  {\bibfnamefont {A.}~\bibnamefont {Fioretti}}, \bibinfo {author}
  {\bibfnamefont {C.}~\bibnamefont {Gabbanini}}, \bibinfo {author}
  {\bibfnamefont {G.}~\bibnamefont {Modugno}}, \bibinfo {author} {\bibfnamefont
  {A.}~\bibnamefont {Recati}},\ and\ \bibinfo {author} {\bibfnamefont
  {S.}~\bibnamefont {Stringari}},\ }\bibfield  {title} {\bibinfo {title}
  {{Supersolid symmetry breaking from compressional oscillations in a dipolar
  quantum gas}},\ }\href {https://doi.org/10.1038/s41586-019-1568-6} {\bibfield
   {journal} {\bibinfo  {journal} {Nature}\ }\textbf {\bibinfo {volume}
  {574}},\ \bibinfo {pages} {382} (\bibinfo {year} {2019})},\ \Eprint
  {https://arxiv.org/abs/1906.02791} {arXiv:1906.02791} \BibitemShut {NoStop}%
\bibitem [{\citenamefont {Guo}\ \emph {et~al.}(2019)\citenamefont {Guo},
  \citenamefont {B{\"{o}}ttcher}, \citenamefont {Hertkorn}, \citenamefont
  {Schmidt}, \citenamefont {Wenzel}, \citenamefont {B{\"{u}}chler},
  \citenamefont {Langen},\ and\ \citenamefont {Pfau}}]{Guo19_supersolid_exp}%
  \BibitemOpen
  \bibfield  {author} {\bibinfo {author} {\bibfnamefont {M.}~\bibnamefont
  {Guo}}, \bibinfo {author} {\bibfnamefont {F.}~\bibnamefont {B{\"{o}}ttcher}},
  \bibinfo {author} {\bibfnamefont {J.}~\bibnamefont {Hertkorn}}, \bibinfo
  {author} {\bibfnamefont {J.~N.}\ \bibnamefont {Schmidt}}, \bibinfo {author}
  {\bibfnamefont {M.}~\bibnamefont {Wenzel}}, \bibinfo {author} {\bibfnamefont
  {H.~P.}\ \bibnamefont {B{\"{u}}chler}}, \bibinfo {author} {\bibfnamefont
  {T.}~\bibnamefont {Langen}},\ and\ \bibinfo {author} {\bibfnamefont
  {T.}~\bibnamefont {Pfau}},\ }\bibfield  {title} {\bibinfo {title} {{The
  low-energy Goldstone mode in a trapped dipolar supersolid}},\ }\href
  {https://doi.org/10.1038/s41586-019-1569-5} {\bibfield  {journal} {\bibinfo
  {journal} {Nature}\ }\textbf {\bibinfo {volume} {574}},\ \bibinfo {pages}
  {386} (\bibinfo {year} {2019})},\ \Eprint {https://arxiv.org/abs/1906.04633}
  {arXiv:1906.04633} \BibitemShut {NoStop}%
\bibitem [{\citenamefont {Natale}\ \emph {et~al.}(2019)\citenamefont {Natale},
  \citenamefont {van Bijnen}, \citenamefont {Patscheider}, \citenamefont
  {Petter}, \citenamefont {Mark}, \citenamefont {Chomaz},\ and\ \citenamefont
  {Ferlaino}}]{Natale2019_supersolid_exp}%
  \BibitemOpen
  \bibfield  {author} {\bibinfo {author} {\bibfnamefont {G.}~\bibnamefont
  {Natale}}, \bibinfo {author} {\bibfnamefont {R.~M.~W.}\ \bibnamefont {van
  Bijnen}}, \bibinfo {author} {\bibfnamefont {A.}~\bibnamefont {Patscheider}},
  \bibinfo {author} {\bibfnamefont {D.}~\bibnamefont {Petter}}, \bibinfo
  {author} {\bibfnamefont {M.~J.}\ \bibnamefont {Mark}}, \bibinfo {author}
  {\bibfnamefont {L.}~\bibnamefont {Chomaz}},\ and\ \bibinfo {author}
  {\bibfnamefont {F.}~\bibnamefont {Ferlaino}},\ }\bibfield  {title} {\bibinfo
  {title} {{Excitation Spectrum of a Trapped Dipolar Supersolid and Its
  Experimental Evidence}},\ }\href
  {https://doi.org/10.1103/PhysRevLett.123.050402} {\bibfield  {journal}
  {\bibinfo  {journal} {Phys. Rev. Lett.}\ }\textbf {\bibinfo {volume} {123}},\
  \bibinfo {pages} {050402} (\bibinfo {year} {2019})},\ \Eprint
  {https://arxiv.org/abs/1907.01986} {arXiv:1907.01986} \BibitemShut {NoStop}%
\bibitem [{\citenamefont {Norcia}\ \emph {et~al.}(2021)\citenamefont {Norcia},
  \citenamefont {Politi}, \citenamefont {Klaus}, \citenamefont {Poli},
  \citenamefont {Sohmen}, \citenamefont {Mark}, \citenamefont {Bisset},
  \citenamefont {Santos},\ and\ \citenamefont
  {Ferlaino}}]{Norcia2021_supersolid_2D}%
  \BibitemOpen
  \bibfield  {author} {\bibinfo {author} {\bibfnamefont {M.~A.}\ \bibnamefont
  {Norcia}}, \bibinfo {author} {\bibfnamefont {C.}~\bibnamefont {Politi}},
  \bibinfo {author} {\bibfnamefont {L.}~\bibnamefont {Klaus}}, \bibinfo
  {author} {\bibfnamefont {E.}~\bibnamefont {Poli}}, \bibinfo {author}
  {\bibfnamefont {M.}~\bibnamefont {Sohmen}}, \bibinfo {author} {\bibfnamefont
  {M.~J.}\ \bibnamefont {Mark}}, \bibinfo {author} {\bibfnamefont {R.~N.}\
  \bibnamefont {Bisset}}, \bibinfo {author} {\bibfnamefont {L.}~\bibnamefont
  {Santos}},\ and\ \bibinfo {author} {\bibfnamefont {F.}~\bibnamefont
  {Ferlaino}},\ }\bibfield  {title} {\bibinfo {title} {{Two-dimensional
  supersolidity in a dipolar quantum gas}},\ }\href
  {https://doi.org/10.1038/s41586-021-03725-7} {\bibfield  {journal} {\bibinfo
  {journal} {Nature}\ }\textbf {\bibinfo {volume} {596}},\ \bibinfo {pages}
  {357} (\bibinfo {year} {2021})},\ \Eprint {https://arxiv.org/abs/2102.05555}
  {arXiv:2102.05555} \BibitemShut {NoStop}%
\bibitem [{\citenamefont {Biagioni}\ \emph {et~al.}(2022)\citenamefont
  {Biagioni}, \citenamefont {Antolini}, \citenamefont {Ala{\~{n}}a},
  \citenamefont {Modugno}, \citenamefont {Fioretti}, \citenamefont {Gabbanini},
  \citenamefont {Tanzi},\ and\ \citenamefont
  {Modugno}}]{Biagioni2022_supersolid_2D}%
  \BibitemOpen
  \bibfield  {author} {\bibinfo {author} {\bibfnamefont {G.}~\bibnamefont
  {Biagioni}}, \bibinfo {author} {\bibfnamefont {N.}~\bibnamefont {Antolini}},
  \bibinfo {author} {\bibfnamefont {A.}~\bibnamefont {Ala{\~{n}}a}}, \bibinfo
  {author} {\bibfnamefont {M.}~\bibnamefont {Modugno}}, \bibinfo {author}
  {\bibfnamefont {A.}~\bibnamefont {Fioretti}}, \bibinfo {author}
  {\bibfnamefont {C.}~\bibnamefont {Gabbanini}}, \bibinfo {author}
  {\bibfnamefont {L.}~\bibnamefont {Tanzi}},\ and\ \bibinfo {author}
  {\bibfnamefont {G.}~\bibnamefont {Modugno}},\ }\bibfield  {title} {\bibinfo
  {title} {{Dimensional Crossover in the Superfluid-Supersolid Quantum Phase
  Transition}},\ }\href {https://doi.org/10.1103/PhysRevX.12.021019} {\bibfield
   {journal} {\bibinfo  {journal} {Phys. Rev. X}\ }\textbf {\bibinfo {volume}
  {12}},\ \bibinfo {pages} {021019} (\bibinfo {year} {2022})},\ \Eprint
  {https://arxiv.org/abs/2111.14541} {arXiv:2111.14541} \BibitemShut {NoStop}%
\bibitem [{\citenamefont {Bloch}\ and\ \citenamefont
  {Greiner}(2022)}]{Bloch22_overview}%
  \BibitemOpen
  \bibfield  {author} {\bibinfo {author} {\bibfnamefont {I.}~\bibnamefont
  {Bloch}}\ and\ \bibinfo {author} {\bibfnamefont {M.}~\bibnamefont
  {Greiner}},\ }\bibfield  {title} {\bibinfo {title} {{The superfluid-to-Mott
  insulator transition and the birth of experimental quantum simulation}},\
  }\href {https://doi.org/10.1038/s42254-022-00520-9} {\bibfield  {journal}
  {\bibinfo  {journal} {Nat. Rev. Phys.}\ }\textbf {\bibinfo {volume} {4}},\
  \bibinfo {pages} {739} (\bibinfo {year} {2022})}\BibitemShut {NoStop}%
\bibitem [{\citenamefont {Esslinger}(2010)}]{Esslinger2010_fermion_QSim_rev}%
  \BibitemOpen
  \bibfield  {author} {\bibinfo {author} {\bibfnamefont {T.}~\bibnamefont
  {Esslinger}},\ }\bibfield  {title} {\bibinfo {title} {{Fermi-Hubbard Physics
  with Atoms in an Optical Lattice}},\ }\href
  {https://doi.org/10.1146/annurev-conmatphys-070909-104059} {\bibfield
  {journal} {\bibinfo  {journal} {Annu. Rev. Condens. Matter Phys.}\ }\textbf
  {\bibinfo {volume} {1}},\ \bibinfo {pages} {129} (\bibinfo {year}
  {2010})}\BibitemShut {NoStop}%
\bibitem [{\citenamefont {Vale}\ and\ \citenamefont
  {Zwierlein}(2021)}]{Vale2021_QSim_rev}%
  \BibitemOpen
  \bibfield  {author} {\bibinfo {author} {\bibfnamefont {C.~J.}\ \bibnamefont
  {Vale}}\ and\ \bibinfo {author} {\bibfnamefont {M.}~\bibnamefont
  {Zwierlein}},\ }\bibfield  {title} {\bibinfo {title} {{Spectroscopic probes
  of quantum gases}},\ }\href {https://doi.org/10.1038/s41567-021-01434-6}
  {\bibfield  {journal} {\bibinfo  {journal} {Nature Physics}\ }\textbf
  {\bibinfo {volume} {17}},\ \bibinfo {pages} {1305} (\bibinfo {year}
  {2021})}\BibitemShut {NoStop}%
\bibitem [{\citenamefont {Towns}\ \emph {et~al.}(2014)\citenamefont {Towns},
  \citenamefont {Cockerill}, \citenamefont {Dahan}, \citenamefont {Foster},
  \citenamefont {Gaither}, \citenamefont {Grimshaw}, \citenamefont {Hazlewood},
  \citenamefont {Lathrop}, \citenamefont {Lifka}, \citenamefont {Peterson},
  \citenamefont {Roskies}, \citenamefont {Scott},\ and\ \citenamefont
  {Wilkins-Diehr}}]{xsede}%
  \BibitemOpen
  \bibfield  {author} {\bibinfo {author} {\bibfnamefont {J.}~\bibnamefont
  {Towns}}, \bibinfo {author} {\bibfnamefont {T.}~\bibnamefont {Cockerill}},
  \bibinfo {author} {\bibfnamefont {M.}~\bibnamefont {Dahan}}, \bibinfo
  {author} {\bibfnamefont {I.}~\bibnamefont {Foster}}, \bibinfo {author}
  {\bibfnamefont {K.}~\bibnamefont {Gaither}}, \bibinfo {author} {\bibfnamefont
  {A.}~\bibnamefont {Grimshaw}}, \bibinfo {author} {\bibfnamefont
  {V.}~\bibnamefont {Hazlewood}}, \bibinfo {author} {\bibfnamefont
  {S.}~\bibnamefont {Lathrop}}, \bibinfo {author} {\bibfnamefont
  {D.}~\bibnamefont {Lifka}}, \bibinfo {author} {\bibfnamefont {G.~D.}\
  \bibnamefont {Peterson}}, \bibinfo {author} {\bibfnamefont {R.}~\bibnamefont
  {Roskies}}, \bibinfo {author} {\bibfnamefont {J.~R.}\ \bibnamefont {Scott}},\
  and\ \bibinfo {author} {\bibfnamefont {N.}~\bibnamefont {Wilkins-Diehr}},\
  }\bibfield  {title} {\bibinfo {title} {{XSEDE: Accelerating Scientific
  Discovery}},\ }\href {https://doi.org/10.1109/MCSE.2014.80} {\bibfield
  {journal} {\bibinfo  {journal} {Comput. Sci. Eng.}\ }\textbf {\bibinfo
  {volume} {16}},\ \bibinfo {pages} {62} (\bibinfo {year} {2014})}\BibitemShut
  {NoStop}%
\end{thebibliography}%
